\documentclass{JHEP3} % 10pt is ignored!

\usepackage{graphicx}
\usepackage{subfigure}
\newcommand\fverb{\setbox\fverbbox=\hbox\bgroup\verb}
\newcommand\fverbdo{\egroup\medskip\noindent%
            \fbox{\unhbox\fverbbox}\ }
\newcommand\fverbit{\egroup\item[\fbox{\unhbox\fverbbox}]}
\newbox\fverbbox

\newcommand{\be}{\begin{equation}}
\newcommand{\ee}{\end{equation}}
\newcommand{\bea}{\begin{eqnarray}}
\newcommand{\eea}{\end{eqnarray}}
\newcommand{\ba}{\begin{array}}
\newcommand{\ea}{\end{array}}

\def\a{\alpha}

\def\o{\omega}

\def\th{\theta}
\def\r{\rho}

\def\t{\tau}

\def\O{\Omega}

\title{Symmetry Energy and Universality classes of  holographic QCD}

\author{Yunseok Seo
\\
Center for Quantum Spacetime, Sogang University, Seoul 121-742, Korea
\\ E-mail: \email{yseo@sogang.ac.kr}}

\author{Sang-Jin Sin
\\
Department of Physics, Hanyang University, Seoul 133-791, Korea
\\ E-mail: \email{sjsin@hanyang.ac.kr}}
\received{\today}       %%
\abstract{
We study nuclear symmetry energy of dense matter using holographic QCD. 
We calculate it in a various holographic QCD models 
 and  show that the scaling index of the symmetry energy 
in dense medium is almost invariant under the smooth  deformation
of the metric as well as  the embedding shape of the probe brane.  
We find that the scaling index depends only on the dimensionality of the branes and space-time. 
Therefore the scaling index of the symmetry energy  characterizes the universality classes  of holographic QCD models.  We suggest  that  the scaling index  might be 
also related to the non-fermi liquid behavior of the interacting nucleons. }


\begin{document}

\section{Introduction}
 
Nuclear symmetry energy is one of key words in nuclear physics as well as  in astrophysics.
Its density dependence is a core quantity of asymmetric nuclear matter which
has important effects on heavy nuclei and is essential to understand neutron star properties.
A big surprise is that such an important quantity is still 
 poorly understood after 80 years of its definition,  
especially in the  supra-saturation density regime. See references \cite{sE1,sE2,sE3,sE5,sE6,sE7,Lee:2010sw,XLCYZ,Che05a} for a review and for a recent discussion.
Not much data is available from experimental side and 
theoretical calculations 
showed  all possible results in high density such that no consensus  could be reached:  some showed
stiff dependence on  density, while others showed soft one depending on models and parameters. 

Given this situation, it would be very interesting if we can  examine the
behavior of the nuclear symmetry energy at high densities
with  a reliable calculational tool.
Recently \cite{Kim:2010dp}  we used a gauge/gravity duality~\cite{Maldacena:1997re, Gubser:1998bc, Witten:1998qj,D4D6_03,hQCD}  to calculate the nuclear symmetry energy. 
We treated the  dense matter in confined phase
 using the method developed in our previous paper \cite{Seo:2008qc,KSS2010}.
Our result showed that the symmetry energy should be stiff in high density 
and its density dependence goes like $\sim \rho^{1/2}$. There, we attributed the stiffness to the repulsion due to the
Pauli principle and suggested the relation of the scaling exponent to the anomalous dispersion relation. 
 
The purpose of this paper is to examine how universal or robust  is the result. 
If the result changes under small variation of the gluon dynamics,  
the result is not very interesting since the true QCD dual is not yet known. 
Only when the result is largely background independent, it can be considered as an interesting one. 
The universality of the $\eta/s$ is the reason why it is interesting even though it is not calculated in the QCD itself. 

We will first calculate the deformation of the metric under certain class of the D brane configurations and show that 
the result is not much dependent on the metric deformation. The scaling behavior is rather insensitive 
whether we use the flat embedding or exact shape of the brane embedding, showing the universality of the 
result. On the other hand, we will see that the scaling exponent depends on the dimensionality of the 
color and flavor branes. We call such discrete dependence of the scaling dimension as the universality class 
of the symmetry energy. 

The rest of the paper is summarized as follows. 
In section 2, we give a definition of symmetry energy and general formula in the brane set up. 
In section 3, we calculate the symmetry energy  $S_2$ in  nuclear matter using D4- as well as D3- 
{\it confining geometry} for  various  probe branes. 
In section 4, we calculate  $S_2$ in  quark  matter using D4, D3 
{\it de-confining geometry}. 
In section 5, we reproduce the scaling exponent analytically using the BPS background and flat embedding
and thereby argue that it is a invariant under the smooth deformation of the gluon dynamics.  
In section 6, we discuss the possible relation of the scaling exponent with the non-fermi liquid 
nature of the strongly interacting dense matter system and conclude. 

%%%%%
\section{Symmetry energy}
Let $N, Z$ be the nucleon and proton numbers respectively, $A$ be their sum 
and $\rho$ is total baryon density. 
 Then energy {\it per nucleon} in 
 nuclear matter system can be expanded as
a function of the isospin asymmetry parameter  $\a=(N-Z)/A$, 
\be\label{E0}
E(\rho,\a)=E(\rho,0) + E_{\rm sym}(\rho)\a^2 +O(\a^4).
\ee
The bulk nuclear symmetry is defined as the coefficient $E_{\rm sym}(\rho)$ in the above expansion. 
There is no term which is odd power in $\a$ due to the exchange symmetry between protons and neutron in nuclear matter.  It is a energy cost per nucleon to deviate the line $Z=N$.

To calculate nuclear symmetry energy in holographic QCD, 
we introduce two flavor $Dq$ branes for  up and down quarks in the metric background created by the $N_c$ of $Dp$ color branes. 
For simplicity, we assume that masses of up and down quarks are the same
so that two branes have same asymptotic positions. 
For the confining geometry,
we can introduce a baryon by the baryon vertex  \cite{Witten:1998xy} 	 which is a compact $D(8-p)$ branes wrapping the $S^{8-p}$ transverse to the $D(8-p)$. If there are $Q$ of them we distribute them homogeneously along the 3 non-compact spatial direction of $Dp$.
From each baryon vertex, $N_c$ of the strings   emanate and end  at one of the 
probe branes. Let $Q_1$, $Q_2$ strings  end on up and down branes respectively.  The end points of the strings have $U(1)$ charges that will create the   $U(1)$ gauge field on each brane. Such $U(1)$ charges are responsible  for the   quark
density of each type of quarks. 

The total free energy of the system can be written as 
\be
{\cal F}_{\rm total}(Q) ={\cal F}(Q) +{\cal F}_{Dq}^{(1)}(Q_1) + {\cal F}_{Dq}^{(2)}(Q_2),
\label{free}\ee
where $Q$ is number density of source and  ${\cal F}(Q)$ is a quantity which depends only on total charge $Q$ as we will discuss later. We can define total charge  density and asymmetry parameter as 
\be
Q=Q_1 +Q_2,~~~~\tilde{\a}=\frac{Q_1 -Q_2}{Q}.
\ee
If we fix the asymptotic value of two probe brane to be same, the total free energy has minimum at $\tilde{\a} =0$\cite{KSS2010}. Then we can  expand  total free energy in 
$\tilde{\a}$;
\be\label{ham03}
{\cal F}_{\rm total}(\tilde{Q}) = E_0 +E_1 \tilde{\a}  + E_2  \tilde{\a}^2 + \cdots.
\ee
The first term,  $E_0 = {\cal F}(Q)+2 {\cal F}_{Dq}\left(\frac{Q}{2}\right)$,
can be identified with the free energy for symmetric matter. The second term in (\ref{ham03}) is zero because (\ref{free}) is symmetric in $Q_1$, $Q_2$.  
 The symmetry energy  is defined from the energy per nucleon and  given by
 \be\label{E2} 
   S_2 =\frac{E_2(Q)}{Q}  = \frac{Q}{4} \cdot \frac{\partial^2{\cal F}_{Dq}^{(1)}(Q_1)}{\partial Q_1^2} \Bigg|_{Q_1 =Q/2} .\ee
\vskip0.4cm
To calculate symmetry energy from D-brane set up, we consider probe $Dq$ brane  spans along $(t,\vec{x}_d, \rho)$ and wraps $S^n$ where $n=q-d-1$. The induced metric on $Dq$ brane can be written in general form;
\be
ds_{Dq}^2 =-G_{tt} dt^2 +G_{xx} d\vec{x}_{d}^2 +G_{\rho\rho}d\rho^2 + G_{\O\O} d\O_{q-d-1}^2.
\ee
To introduce number density, we turn on time component of $U(1)$ gauge field on the probe brane whose action is given by 
\be\label{DBI01}
S_{Dq} = \mu_q \int d\sigma^{q+1} e^{-\phi}\sqrt{{\rm det}(g + 2\pi \a' F)},
\ee
where $\mu_q$ is tension of $Dq$ brane. The free energy can be identified with the action for fixed charge sector, which is the   Legendre transformation of the original action with respect to the  gauge field,
\be\label{FDq}
{\cal F}_{Dq}(\tilde{Q}) =\tau_q \int d\rho \sqrt{G_{tt}G_{\rho\rho}}\sqrt{\tilde{Q}^2  +e^{-2\phi} G_{xx}^d G_{\O\O}^{n}},
\ee
where $\tau_q =\mu_q V_d \O_n$, $\tilde{Q} =Q/(2\pi \a' \tau_q)$ and $n=q-d-1$.
From (\ref{E2})   we    symmetry energy is given by; 
\be\label{symE}
S_2
= 2\tau_q \int d\r \frac{\tilde{Q}\sqrt{G_{tt}G_{\r\r}}e^{-2\phi}G_{xx}^d G_{\O\O}^{n}}{\left(\tilde{Q}^2 +4 e^{-2\phi}G_{xx}^d G_{\O\O}^{n}\right)^{3/2}}.
\ee
For details, see Appendix \ref{AppA}. 
%Each metric component in (\ref{symE}) is the solution  with fixed charge  and quark mass. 
This result can be applied  for general $Dp$ brane background and 
 we will apply it to  confined phase as well as  deconfined one. 
 
\section{Symmetry energy of nuclear matter}

In this section, we will discuss symmetry energy in the confined phase. 
%We call corresponding geometry as confining geometry. 
There are  several examples of the metric background corresponding to the confining phase.  
In this paper, we will consider two examples  based on $D4$ and $D3$ branes.
 In the case of $D4$ brane, the geometry is obtained by (double) Wick rotating  time and  a compact spatial direction.
  In $D3$ brane case, 
  we will use the non-supersymmetric geometry with nontrivial dilaton field \cite{Gubser:1999pk}.
In such geometries,  the net force on the probe brane is repulsive. 

We identify the the chiral symmetry as the 
the symmetry  rotating the  probe brane in the   transverse plane  following \cite{EHS}.
 In the limit of zero quark mass, this symmetry  is spontaneously broken due to the repulsive nature of the net force. 
As a consequence,  the value of chiral condensation has finite value and 
 the baryon vertex  is allowed to exist. 
 %    wrapping transverse sphere of original $N_c$ brane. Due to the Chern-Simons interaction between background Ramond-Ramond field of $N_c$ brane and world volume of spherical brane, $N_c$ sources should be induced on the spherical brane. We call this configuration as `baryon vertex' \cite{Witten:1998xy} because it is bound state of $N_c$ quark from the boundary theory point of veiw. 
The baryon vertices play a role of source of $U(1)$ gauge field on probe $Dq$ brane as we discussed in previous section. 
% We can connect the end point fundamental strings on baryon vertex to the probe $D_q$ brane. Then, the total configuration is probe brane which carries flavor degrees of freedom with baryon vertices. It can be understood as dense nuclear system at the boundary theory. 

\begin{figure}[!ht]
\begin{center}
\subfigure[]{\includegraphics[angle=0, width=0.47\textwidth]{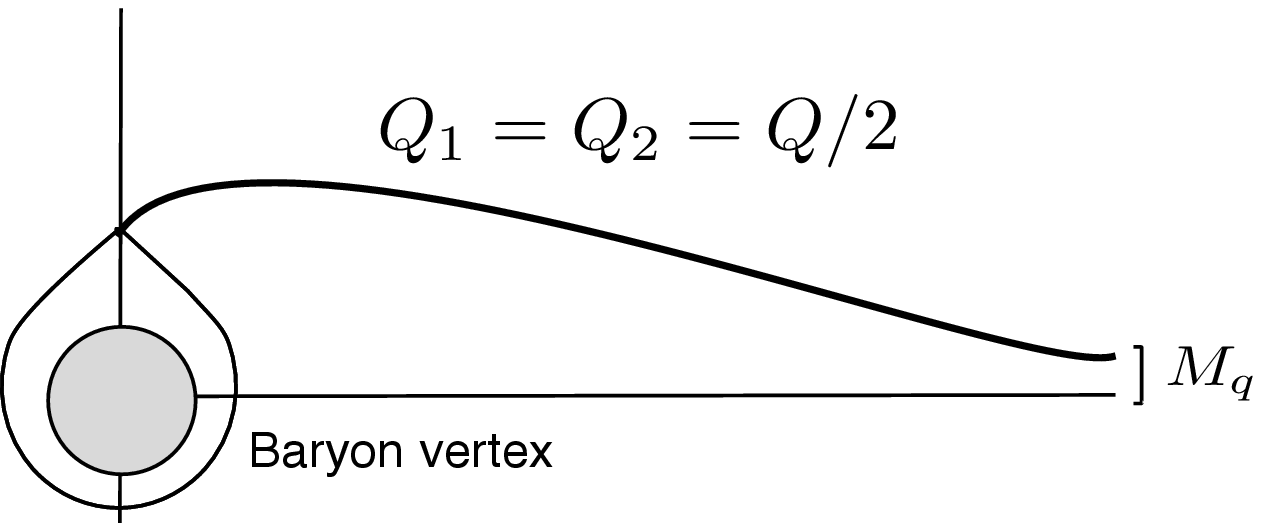}}
\subfigure[]{\includegraphics[angle=0, width=0.47\textwidth]{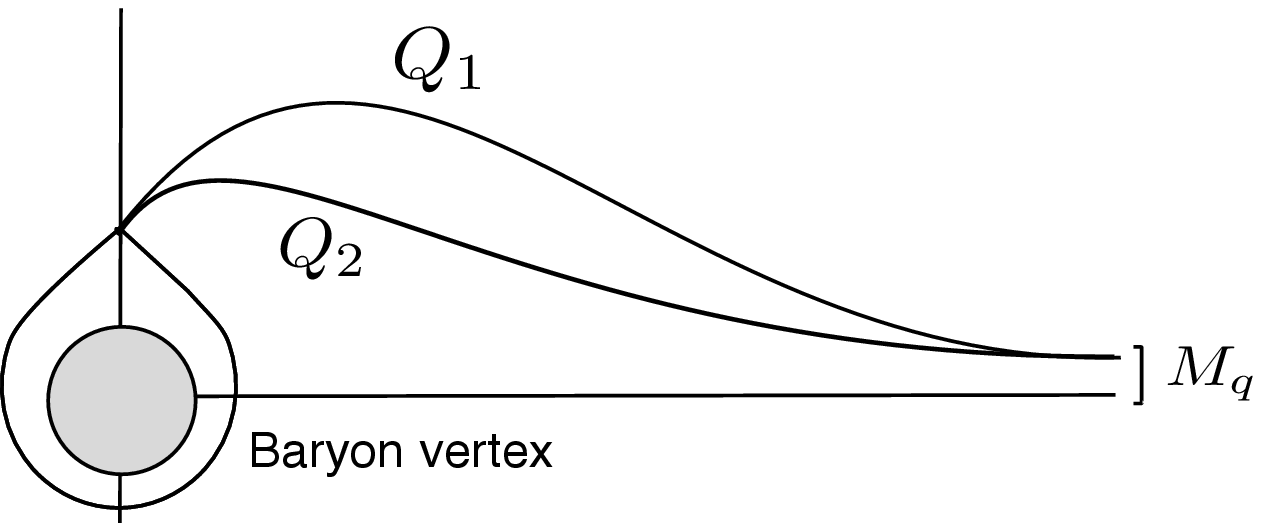}}
\caption{Schematic figure of D-brane configuration for (a) symmetric matter and (b) asymmetric matter. \label{fig:Sconf}}
\end{center}
\end{figure}

To make the system stationary,   we have to impose `force balance condition' between baryon vertex and probe brane. The details of the solution of baryon vertex and force balance condition are in Appendix \ref{AppB}. 
The symmetry energy can be understood as the energy costs when the system is deviated from symmetric matter.  To achieve the deviation, we need to attach  different number of charges(strings)  on each brane, which in turn gives 
different embedding for each probe brane. The symmetry energy (\ref{symE}) is nothing but the energy difference of the D-brane system between symmetric and asymmetric distribution of source on probe branes. The schematic figure is drawn in Figure. \ref{fig:Sconf}. 
We first solve the equation of motion for probe brane numerically with given charge and quark mass together with  force balance condition.  Then substituting the solution  to (\ref{symE}), we can calculate the symmetry energy.

\subsection{$D4$ brane background}
 Here we consider  probe $D6$, $D4$ and $D2$  branes in  the $D4$   background which is given by 
\begin{eqnarray}
ds_{D4}^{2}
&=&\left(\frac{U }{R }\right)^{3/2}\left(\eta_{\mu\nu}dx^\mu dx^\nu + f(U) dx_4^{2} \right)
+\left(\frac{R}{U }\right)^{3/2}\left( \frac{dU^2}{ f(U)} +U^2 d\Omega_4^2\right) \cr
e^\phi&=&g_s\left(\frac{U }{R }\right)^{3/4},\quad F_4 =\frac{2\pi N_c}{\Omega_4}\epsilon_4, \;\; f(U)=
1-\Big(\frac{U_{KK}}{U}\Big)^{3}, \;\; R^3=\pi g_s N_c l_s^3.
\label{adsm}
\end{eqnarray}
The Kaluza-Klein mass scale $M_{KK}$ is defined as inverse radius of the $x_4$ direction:
$M_{KK}=\frac{3}{2}\frac{U^{1/2}_{KK}}{R^{3/2}}$.
The bulk parameters, $U_{KK}, g_s, R$ and the   gauge theory parameters
$M_{KK}, g^2_{YM}, \lambda:=g_{YM}^{2}N_{c}$ are related by
\be\label{consts1}
g_s=\frac{\lambda}{2\pi l_sN_c M_{KK}}, \quad U_{KK}=\frac{2}{9}\lambda M_{KK} l_s^2,
\quad R^3=\frac{\lambda l_s^2}{2M_{KK}} .
\ee
%To make transverse space to D4 brane be flat (up to overall factor), 
We introduce new coordinate $\bar{\xi}$ by 
$\frac{d\bar{\xi}^2}{\bar{\xi}^2}=\frac{dU^2}{U^2f(U)}$
and obtain, in Euclidean signature,
\be\label{d4bgmetric}
ds_{D4}^2 = \left(\frac{U }{R }\right)^{3/2}\left(dt^2 +d\vec{x}^2 + f(U) dx_4^{2} \right)
+\left(\frac{R}{U }\right)^{3/2}\left(\frac{U}{\bar{\xi}}\right)^2\left(d\bar{\xi}^2 +\bar{\xi}^2 d\Omega_4^2\right)\, .
\ee
The relation between $U$ and $\bar\xi$ is
\be \label{uxi}
U^{3/2}=\bar{\xi}^{3/2}\left[1+\left(\frac{\bar{\xi}_{KK}}{\bar{\xi}}\right)^{3}\right]\equiv\bar{\xi}^{3/2}\omega_{+},\ee
where $U_{KK}^{3/2} =2 \bar{\xi}_{KK}^{3/2}$. 
We rescale  $ \bar{\xi} =  \bar{\xi}_{KK}  \xi  $  such that singularity is located at $\xi=1$.
 Then  the background (\ref{d4bgmetric}) and the dilaton can be rewritten as
\bea\label{scaledmet}
ds_{D4}^2 &=&\left(\frac{\bar{\xi}_{KK}}{R}\right)^{3/2} \xi^{3/2} \omega_{+}\left(dt^2 +d\vec{x}^2+f(U)dx_4^2\right) +\left(R^3 \bar{\xi}_{KK}\right)^{1/2} \frac{\omega_{+}^{1/3}}{{\xi}^{3/2}}\left(d{\xi}^2 +{\xi}^2 d\Omega_{4}^2\right),\cr
 e^{\phi}&=&\left(\frac{\bar{\xi}_{KK}}{R}\right)^{3/4}{\xi}^{3/4}\omega_{+}^{1/2}.
\eea
The baryon vertex in this background is a spherical $D4$ brane wrapping $S^4$. 
The induced metric on compact $D4$ brane  is given by   (see \ref{induceBV} for the detail),
\be\label{d4met}
ds_{BV}^2 =\left(\frac{\bar{\xi}_{KK}}{R}\right)^{3/2} {\xi}^{3/2} \omega_{+} dt^2+\left(R^3 \bar{\xi}_{KK}\right)^{1/2} \omega_{+}^{1/3}{\xi}^{1/2}\left[\left(1+\frac{{\xi^\prime}^2}{\xi^2}\right)d\theta^2+\sin^2\theta d\Omega_{3}^2\right].
\ee
The free energy of the compact $D4$ can be written as  
\bea\label{d4h}
{\cal F}_{BV} =\t_4 \int d\theta\sqrt{\omega_+^{4/3} (\xi^2 +\xi'^2)}\sqrt{\tilde{D}(\theta)^2+ \sin^6\theta},
\eea
with
\be\label{displacement}
\tilde{D}(\theta)=-2+3(\cos\theta -\frac{1}{3}\cos^3\theta).
\ee
The embeddings of baryon $D4$ brane can be obtained as a function of $\xi_0$, the position of $D4$ brane at $\xi =0$.
From the embedding solution, we can get value of $\xi$ and its slope at $\th =\pi$ which will be used in  force balance condition. The details of the embeddings are discussed in \cite{Seo:2008qc}.\par

Now we add probe brane in this geometry;  we can put $D6$, $D4$ and $D2$ brane   corresponding to the $3+1$, $2+1$ and $1+1$ theory at the boundary. The general induced metric on $Dq (q=6, 4, 2)$ brane can be written from (\ref{inducedg})
\bea\label{inducedDq}
ds_{D_q}^2&\!=&\!\left(\frac{\bar{\xi}_{KK} }{R }\right)^{3/2}\xi^{3/2}\omega_{+}\left(dt^2 +d\vec{x}_d^2  \right) \cr
&&~~~~~~~~~~+\left(R^3 \bar{\xi}_{KK}\right)^{1/2}\frac{\omega_{+}^{1/3}}{\xi^{3/2}}\left[\left(1+\dot{Y}^2\right)d\r^2 +\r^2 d\Omega_{q-d-1}^2\right].
\eea
From (\ref{ham02}), we can write the free energy of $Dq$ brane as follows
\bea\label{hamDq}
{\cal F}_{Dq} = \hat{\t}_q \int d\rho\sqrt{\o_+^{4/3} (1+\dot{Y}^2)} \sqrt{\hat{Q}^2 + \rho^{2n}\omega_+^{\frac{4}{3}(d-1)}},
\eea
where $n=q-d-1$ and $\hat{\tau}_q \equiv \tau_q \bar{\xi}_{KK}^d$. $\hat{Q}$ is related to the number of source $Q$ as follows
\be
\hat{Q} =\frac{Q}{\t_q (2\pi \a')\bar{\xi}_{KK}^{(d-1)}}.
\ee
 The  boundary condition  is described in the appendix B (\ref{bc}). 
 We  now can solve the equation of motion  with  force balance condition for given $\hat{Q}$. \par
\vskip0.2cm

\begin{figure}[!ht]
\begin{center}
\subfigure[]{\includegraphics[angle=0, width=0.45\textwidth]{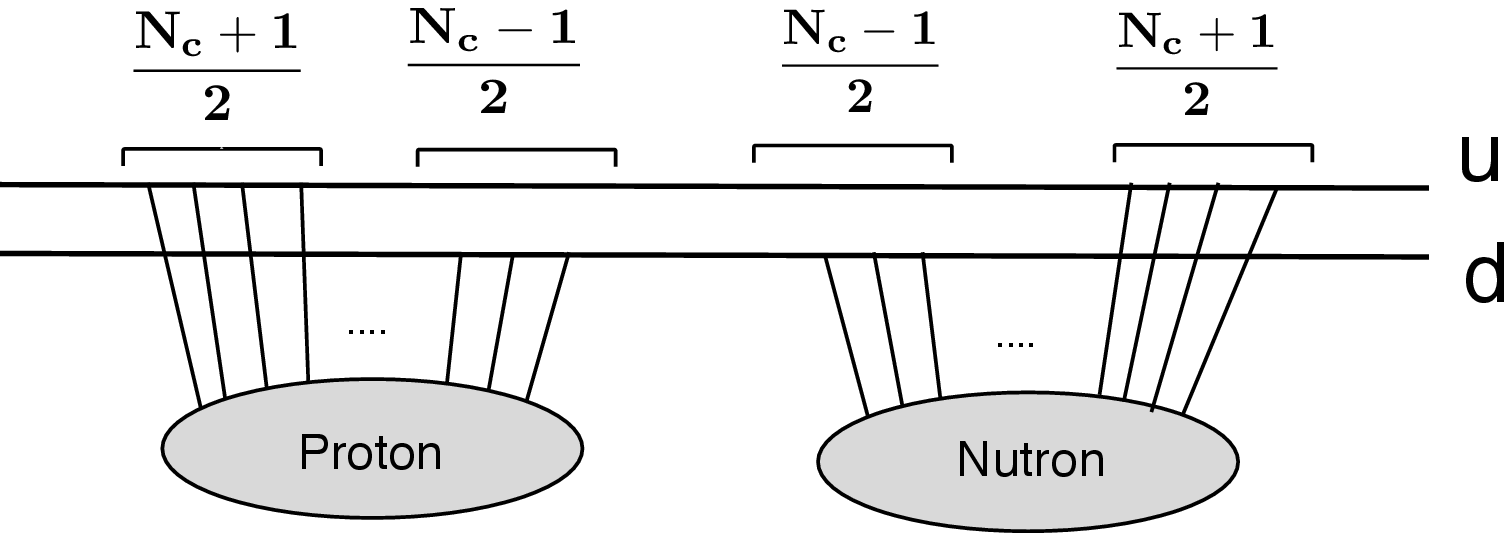}}
~~~~
\subfigure[]{\includegraphics[angle=0, width=0.45\textwidth]{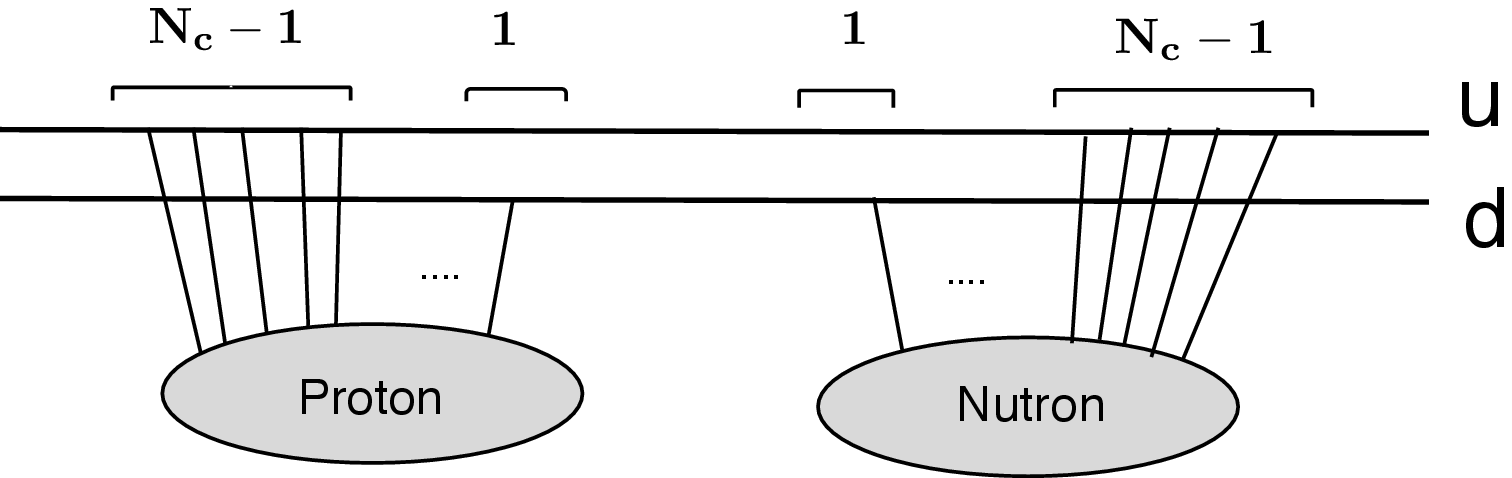}}
\caption{  proton and neutron in generic $N_c$. \label{fig:PN}}
\end{center}
\end{figure}

For the nuclear symmetry energy (\ref{symE}) in large $N_c$ theory, we need to define 
what is the proton and neutron. In the case of $N_c =3$, proton consists of two up quark and one down quark({\it uud}), and neutron is {\it udd}. But in generic $N_c$, it is not clear what is the quark contents of proton and neutron. There are many possibilities for quark configurations of proton and neutron but 
we consider  two possibilities in Figure \ref{fig:PN}.

$\bullet$ case (a): 
In this case, the difference of quark number between proton and neutron is always $1$. To make this configuration be possible, we assume that $N_c$ is odd. From this configuration, we can set
\be
Q_1-Q_2 =N_p -N_n, ~~~~~Q_1 +Q_2 =Q =N_B N_c,
\ee
where $N_p$ is number of proton and $N_n$ is number of neutron. Then, $\tilde{\a}$ can be written as
\be
\tilde{\a} =\frac{Q_1 -Q_2}{Q_1 +Q_2} =\frac{N_p -N_n}{N_c N_B},
\ee
and
\be
Q_1 =\frac{1+\tilde{\a}}{2} N_c N_B,~~~~~Q_2 =\frac{1-\tilde{\a}}{2} N_c N_B.
\ee
From this definition,  the second order term becomes
\be
\tilde{\a}^2 E_2 = \left(\frac{N_p -N_n}{N_c N_B}\right)^2 E_2 =\left(\frac{N_p -N_n}{N_B}\right)^2 \cdot \frac{E_2}{N_c^2}.
\ee
Then, the symmetry energy  per nucleon can be identified as
\be
S_2 = \frac{E_2}{N_c^2 N_B}.
\ee
From (\ref{E2}), we can get symmetry energy per nucleon in terms of elements of induced metric,
\be\label{s2d4}
S_2=\frac{ \bar{\xi}_{KK}}{\pi \a'}\cdot \frac{1}{N_C}\int d\rho \frac{\hat{Q} \sqrt{\o_+^{4/3} (1+\dot{Y}^2)}\,\, \rho^{2n} \o_+^{\frac{4}{3}(d-1)}}{\left(\hat{Q}^2 +4 \rho^{2n} \o_+^{\frac{4}{3}(d-1)}\right)^{3/2}}.
\ee
There is $1/N_C$ factor in the symmetry energy (\ref{s2d4}) which implies that the symmetry energy is suppressed by $N_C$. It is consistent with the definition of proton and neutron: there is only one quark difference between proton and neutron and therefore, for large $N_C$, it is not easy to distinguish these two particle and hence symmetry energy becomes zero for large $N_C$. 
\vskip0.1cm
$\bullet$ case (b): In this case, proton consist of $N_C -1$ up quarks and one down quark, and neutron has single up quark and $N_C -1$ down quark. The total difference and total number can be written in term of proton and neutron number as follows
\be
Q_1 -Q_2 =(N_C -2)(N_p -N_n),~~~~~Q_1 +Q_2 =N_C N_B.
\ee
Then,
\be
\tilde{\a} =\frac{Q_1 -Q_2}{Q_1+Q_2} =\frac{(N_C -2)(N_p -N_n)}{N_C N_B},
\ee
and
\be
\tilde{\a}^2 E_2 = \left(\frac{N_p -N_n}{N_B}\right)^2 \cdot \frac{(N_C -2)^2}{N_C^2}E_2.
\ee
Performing same procedure with previous case, we get symmetry energy as follows;
\be\label{S2D4_2}
S_2=\frac{ \bar{\xi}_{KK}}{\pi \a'}\cdot \frac{(N_C-2)^2}{N_C}\int d\rho \frac{\hat{Q} \sqrt{\o_+^{4/3} (1+\dot{Y}^2)}\,\, \rho^{2n} \o_+^{\frac{4}{3}(d-1)}}{\left(\hat{Q}^2 +4 \rho^{2n} \o_+^{\frac{4}{3}(d-1)}\right)^{3/2}}.
\ee
Instead of suppression by $N_C$, the symmetry energy grows with $N_C$ factor for large $N_C$ limit. 

\begin{figure}[!ht]
\begin{center}
\subfigure[]{\includegraphics[angle=0, width=0.45\textwidth]{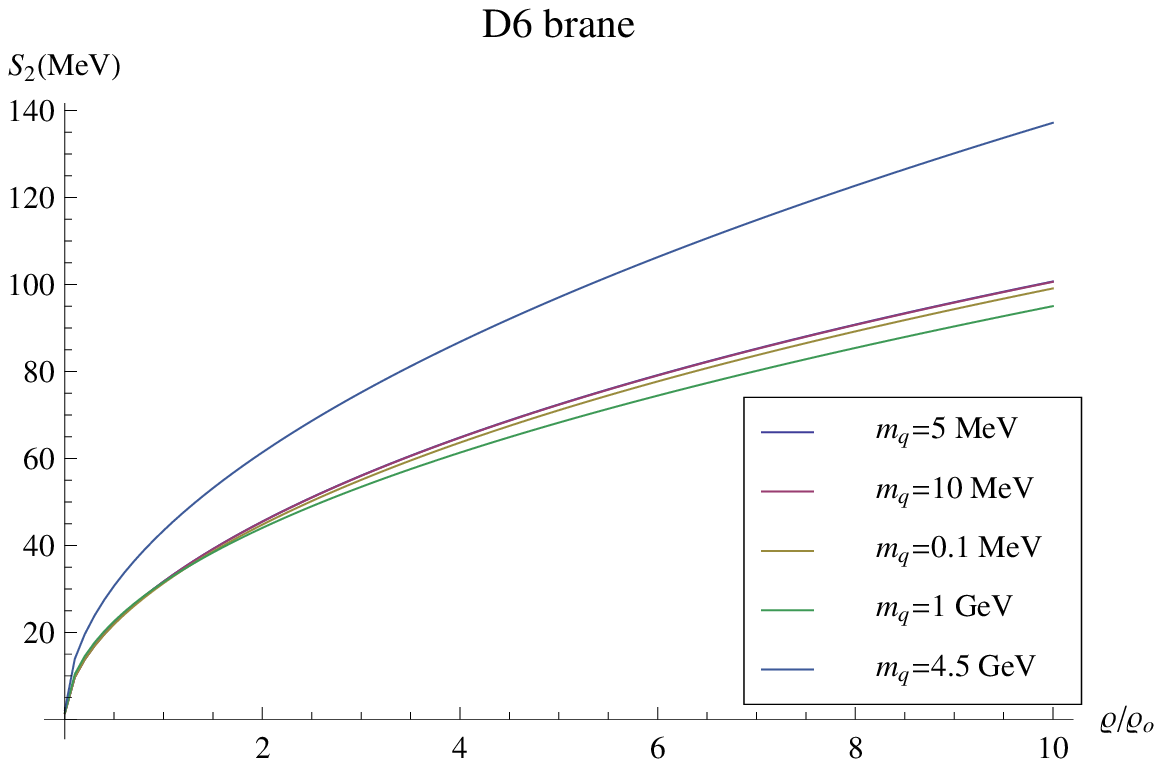}}
~~~~
\subfigure[]{\includegraphics[angle=0, width=0.45\textwidth]{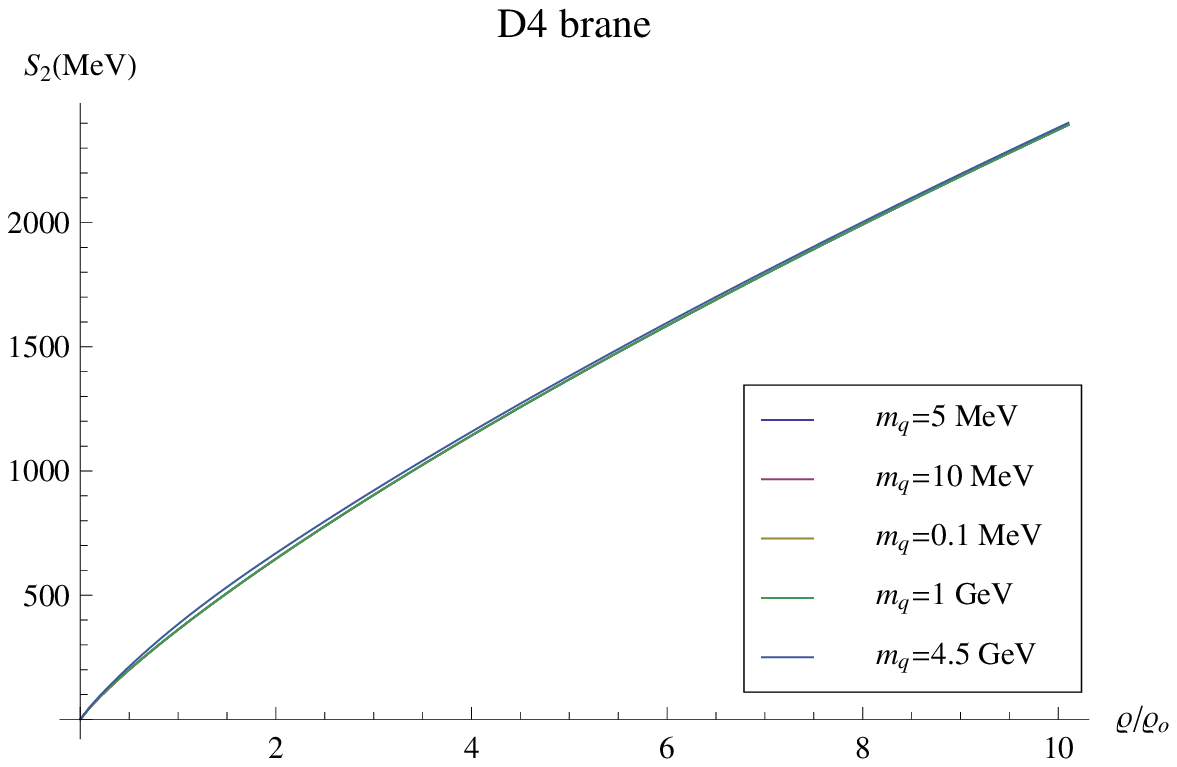}}
\subfigure[]{\includegraphics[angle=0,width=0.45\textwidth]{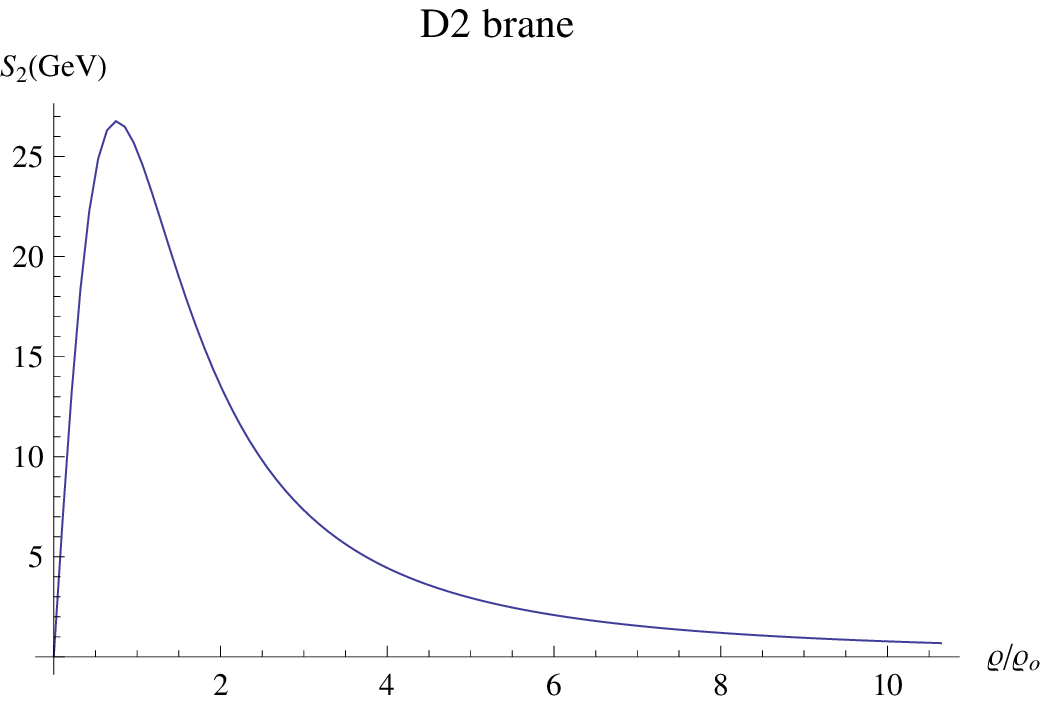}}
\caption{Density dependence of symmetry energy for several quark mass for (a) D6 probe brane, (b) D4 probe brane and (c) D2 probe brane. \label{fig:S2_D4}}
\end{center}
\end{figure}

Considering other intermediate case in similar fashion, we can easily see that 
the free energy should be lowest in the case (a). Therefore we take the definition of 
proton defined in (a).  
%
%But it is also natural for the definition of proton and neutron. In this case, for large $N_C$, when proton is changed to neutron, there is $N_C$ number of quarks are changed. Therefore, symmetry energy should be proportional to $N_C$. In the case of $N_C =3$, two forms have same factor $1/3$.
\vskip0.2cm
 We can convert all parameters in terms of physical quantities such as 't Hooft coupling $\lambda$, Kaluza-Klein scale $M_{KK}$ and density. The nuclear density and quark mass can be written as
\bea
\varrho &=& \frac{Q}{N_c V_d} = \frac{2^{(d-1)/3} \Omega_n}{3^{2(d-1)} (2\pi)^{n+d-1}} \lambda^{d-2} M_{KK}^d \hat{Q},
\cr\cr
m_q &=& \frac{\lambda M_{KK} Y_{\infty}}{2^{2/3} \cdot 9\pi},
\eea
and the coefficient in (\ref{s2d4}) becomes
$\frac{ \bar{\xi}_{KK}}{\pi \a'} =\frac{2^{1/3}}{9\pi} \lambda M_{KK}.$
 By substituting these numerical solution into (\ref{symE}) we can get symmetry energy for each embeddings in terms of density and quark mass. From the meson mass calculation, we choose 
 \be 
 \lambda =18, \;\; M_{KK} =1.04  GeV, \;\;  N_C=3. \ee
  With this values we can calculate density and quark mass dependence of symmetry energy for $D6$, $D4$ and $D2$ probe brane cases. See Figure \ref{fig:S2_D4}.

$\bullet$ $D6$ probe brane:\\
In this case, the   boundary theory is in $3+1$ dimension. 
From the Figure \ref{fig:S2_D4} (a), the symmetry energy with D6 probe brane seems to have square root behavior.   All the lines are well fitted to $S_2 =S_0 (\varrho/\varrho_0)^{1/2}$, with  $27 {\rm MeV} \le S_0 \le 36 {\rm MeV}$. For small quark mass($5 {\rm MeV}$), symmetry energy curve is fitted to $S_2 \sim 28 ({\rm MeV}) (\varrho/\varrho_0)^{1/2}$. As quark mass increase, the symmetry energy curve move downwards, in other words, it become softer up to quark mass is around 100 MeV. After then, the symmetry energy curve moves to upwards(stiffer) as quark mass increases.

$\bullet$ $D4$ probe brane:\\
The dual  theory  lives in 2+1 dimension and the result  is drawn in Figure \ref{fig:S2_D4} (b). In this figure, most symmetry energy curves  are on top of each other unless  quark mass is very large. And it's behavior seems to be linear for small density. 

$\bullet$ $D2$ probe brane:\\
In this case,   dual theory is of 1+1 dimensional. 
The density dependence  in the free energy (\ref{hamDq}) factors  out: 
\be
{\cal F}_{D_2} = \hat{\t}_2 \int d\rho \sqrt{\hat{Q}^2 +1} \sqrt{\o_+^{4/3} (1+\dot{Y}^2)},
\ee
so that the embedding configuration is independent of density.  
The symmetry energy can be written as follows
\be
 S_2=\frac{ \bar{\xi}_{KK}}{4\pi \a'} \frac{\hat{Q}}{(\hat{Q}^2 +4)^{3/2}} \int d\rho \sqrt{\o_+^{4/3} (1+\dot{Y}^2)}.
 \ee
The embedding  configuration looks flat for any quark mass which means that the symmetry energy is almost same for all quark mass region. See Figure. \ref{fig:S2_D4} (c).

\subsection{$D3$ brane background}
We now consider the  confining geometry based on $D3$ brane \cite{DFG};
\be\label{ggmetric}
ds_{10}^2 =e^{\phi/2}\left(\frac{r^2}{R^2} A^2(r)\eta_{\mu\nu} dx^{\mu}dx^{\nu}
+\frac{R^2}{r^2} dr^2 +R^2 d\Omega_5^2 \right),
\ee
where $A(r)$ and the dilaton are given by:
\be
A(r)=\left(1-\left(\frac{r_{0}}{r}\right)^8 \right)^{1/4}  \qquad \& \qquad 
e^{\phi}=\left(\frac{(r/r_0)^4+1}{(r/r_0)^4-1}\right)^{\sqrt{3/2}} \, ,
\ee
while the five-form remains unaltered from the pure $AdS$ solution. 
$R$ is the $AdS$ radius and $r_0$ is the position of the singularity determined by  the value of gluon condensation. Unlike  D4 brane background, the exact relation between $r_0$ and $<G^2>$ is not clear.  So we will see power behavior of symmetry energy in terms of density and asymptotic value of probe brane. For simplicity, we set $r_0 =1$.  Procedding similarly with the previous section, we can get Hamiltonian for baryon $D5$ brane;
\bea\label{hamil-D5}
{\cal F}_{BV}= \t_5 \int d\theta A(r)\sqrt{e^{\phi}(r'^2 +r^2)}\sqrt{D(\theta)^2 +\sin^8\theta}\, ,
\eea
where,
\be
D(\theta)=-\frac{3}{2}\theta+\frac{3}{2}\sin\theta\cos\theta +\sin^3\theta\cos\theta\, .
\ee
The details of this embedding are discussed in \cite{SSSZ}. We add $D7$, (or $D5$, $D3$) brane as a probe. The induced metric on $Dq$ brane can be written as
\be\label{inducedDq2}
ds_{Dq}^2 =e^{\phi/2}\left[\frac{r^2}{R^2} A^2(r)(dt^2+ d\vec{x}_{d}^{2})
+\frac{R^2}{r^2}\left\{(1+\dot{Y}^2 )d\rho^2 + \rho^2 d\Omega_{q-d-1}^2 \right\} \right].
\ee
And the free energy for probe brane is
\bea\label{hamil-D7}
{\cal F}_{Dq} &=& \t_q \int d\rho  \sqrt{e^{\phi}A(r)^2 (1+\dot{Y}^2)} \sqrt{\tilde{Q}^2 + e^{\frac{\phi}{2}(q-5)} A(r)^{2d} \rho^{2 n} }\, .
\eea
The symmetry energy(\ref{symE}) becomes
\be
S_2 =\frac{1}{4\pi \a'}\int d\rho \frac{\tilde{Q}\rho^{2n} e^{\frac{\phi}{2}(q-4)}A(r)^{2d+1}\sqrt{1+\dot{Y}^2}}{\left(\tilde{Q}^2 +4 e^{\frac{\phi}{2}(q-5)} A(r)^{2d} \rho^{2 n}\right)^{3/2} }.
\ee

\begin{figure}[!ht]
\begin{center}
\subfigure[]{\includegraphics[angle=0, width=0.45\textwidth]{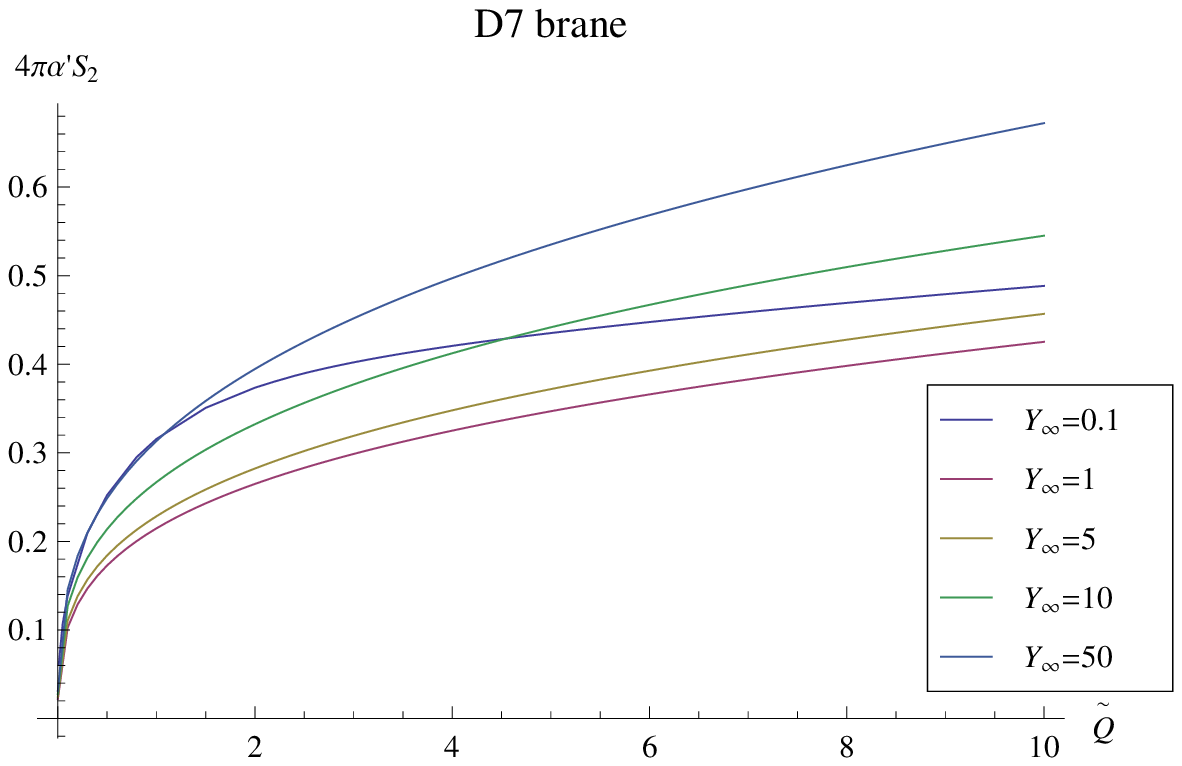}}
~~~~~
\subfigure[]{\includegraphics[angle=0, width=0.45\textwidth]{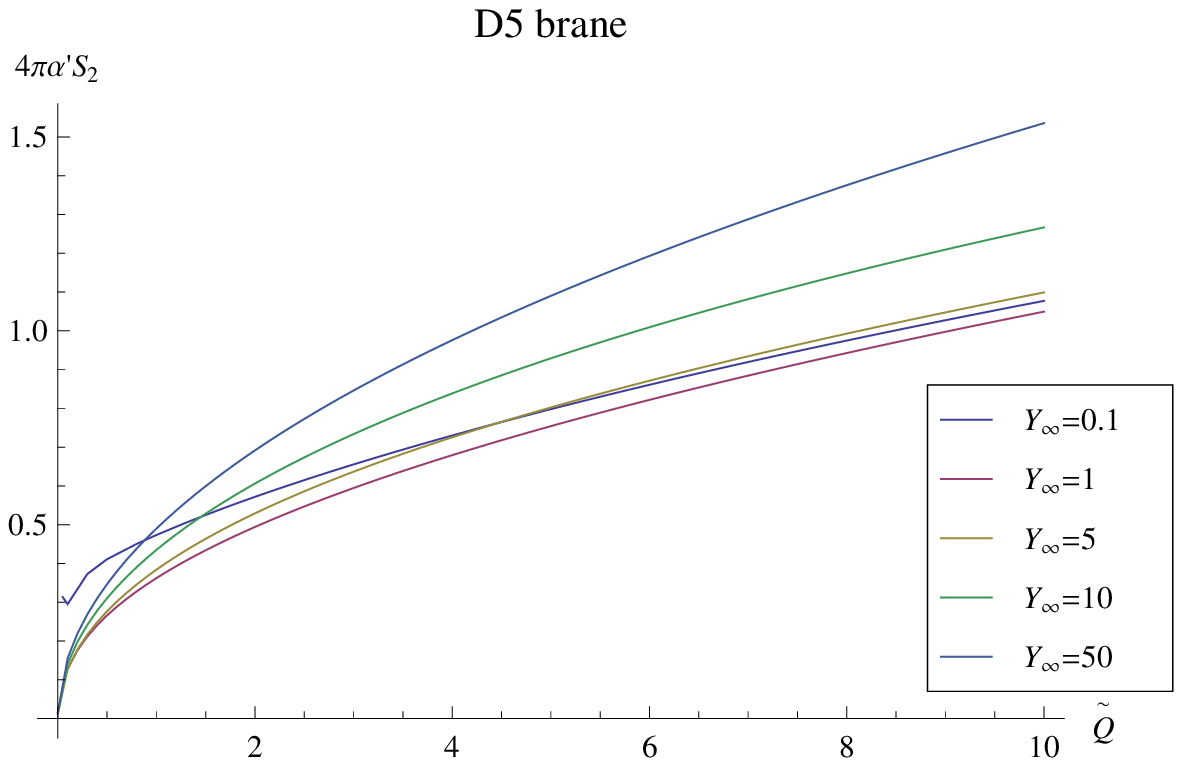}}
\subfigure[]{\includegraphics[angle=0, width=0.45\textwidth]{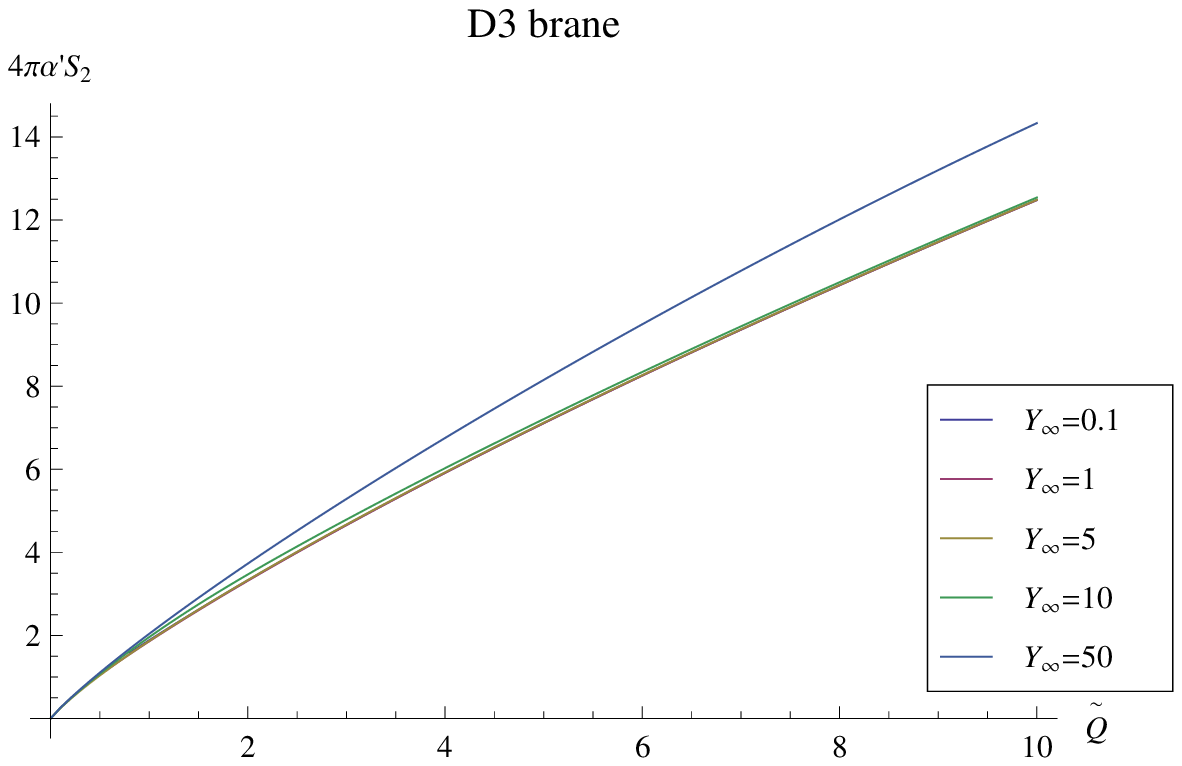}}
\caption{Density dependence of symmetry energy for several quark mass. \label{fig:S2_D3}}
\end{center}
\end{figure}

With same method of the  previous section, we can calculate symmetry energy, see Figure \ref{fig:S2_D3}.

$\bullet$ $D7$ probe brane:\\
In this case, the corresponding boundary boundary theory is of 3+1 dimension. 
The density and $Y_{\infty}$ dependence of symmetry energy is drawn in Figure{\ref{fig:S2_D3}. In this figure, one can see that symmetry energy has power behavior unless   asymptotic value of probe brane($Y_{\infty}=0.1$) is very small.  Actually,  the symmetry energy curves with $1 \le Y_{\infty} \le 50$  
behave as   
\be  S_2 \simeq  \frac{S_0}{4\pi \a'}  \tilde{Q}^{1/3},  
\ee 
with   $0.21 \le   S_0  \le 0.31 $.
\vskip .1cm 

$\bullet$ $D5$ probe brane:\\
 The boundary theory is of 2+1 dimension and the symmetry energy  
 behaves 
 \be  S_2 \simeq  \frac{S_0}{4\pi \a'}  \tilde{Q}^{1/2},  
\ee 
with   $0.34 \le   S_0  \le 0.5 $  unless $Y_{\infty}  $ is very small,  see Figure \ref{fig:S2_D3} (b).
\vskip .1cm

$\bullet$ $D3$ probe brane:\\
The boundary theory is of   1+1 dimension. 
 We can see the symmetry energy grows linearly for small density, see Figure \ref{fig:S2_D3} (c).
\vskip0.2cm

For the given boundary space-time, 
the density dependence of symmetry energy  seems to  depend on  the dimensionality  of D-brane system we use. 
For the 3+1 dimensional boundary theory,  the probe brane should be  $D6$ for $D4$ brane background and  
it should be $D7$ for $D3$ background.  In the case of $D6$ brane,   $S_2 \sim \varrho^{1/2}$. 
On the other hand  $S_2 \sim \varrho^{1/3}$ for $D7$ probe in $D3$ background. 
These differences appear in all other probe brane cases as well. We will discuss these phenomena further  later.

\section{Symmetry energy in quark matter system}
So far, we discuss symmetry energy in confined system.
In this section we  consider symmetry energy in deconfined phase, namely, in  quark matter  instead of nuclear matter. The background geometry  is a black brane. To introduce finite density or chemical potential, we  introduce  fundamental strings which connect black hole horizon and probe brane. The end points of fundamental strings attached on probe  provide sources of $U(1)$ gauge field. Since the fundamental strings can move freely along the $D3$ direction of the black brane horizon, the boundary system can be identified as the system of freely moving quarks. 

Similarly to the previous section, the symmetry energy can be understood  as the energy cost to separate the number of up and down quarks from symmetric matter. From the D-brane point of view, we need to consider two probe branes, 
where different number of strings are attached so that the embedding of  two branes are separated from each other.  The schematic figure is drawn in Figure \ref{fig:Sdeconf}. 

\begin{figure}[!ht]
\begin{center}
\subfigure[]{\includegraphics[angle=0, width=0.47\textwidth]{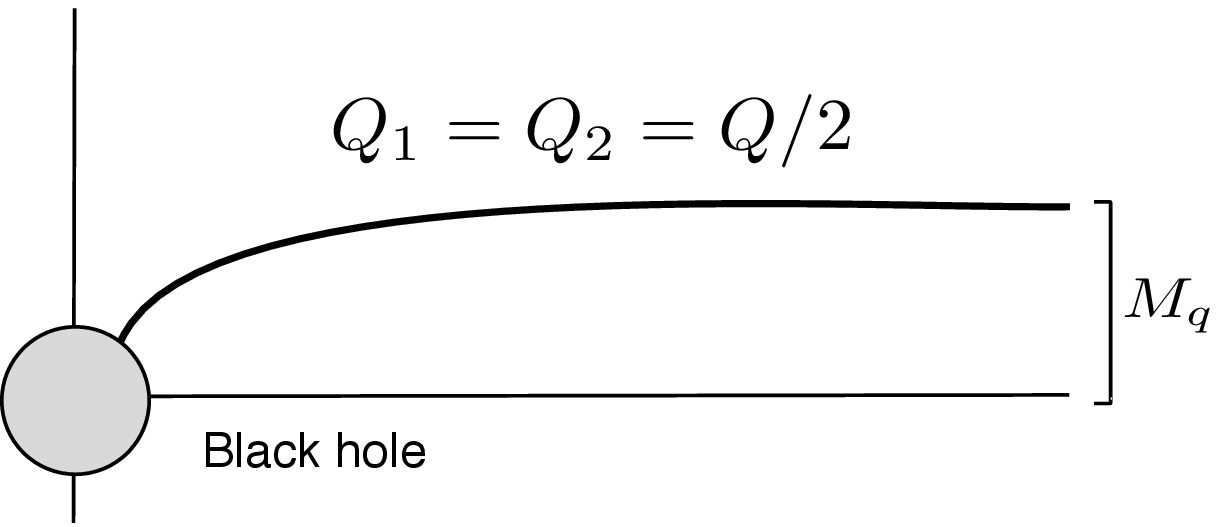}}
~~~~
\subfigure[]{\includegraphics[angle=0, width=0.47\textwidth]{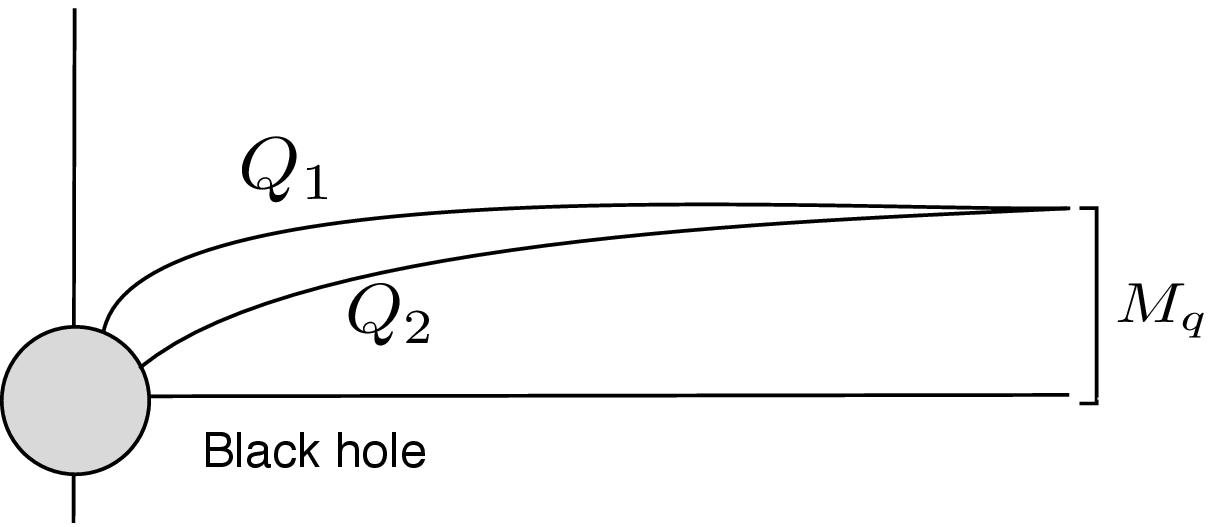}}
\caption{Schematic figure of D-brane configuration for (a) symmetric matter and (b) asymmetric matter. \label{fig:Sdeconf}}
\end{center}
\end{figure}

The  boundary condition at the horizon  is determined by the regularity of the equation of motion at the   horizon;
\be\label{bc2}
\dot{Y}_{\rho_{min}} = \tan\th,
\ee
where $\th$ is the angle of probe brane at the horizon. By substituting the embedding solution into (\ref{symE}), we can get symmetry energy as a function of density and quark mass.

\subsection{$D4$ brane background}
To describe deconfined phase at the boundary theory, we introduce black hole geometry based on D4 brane;
\bea\label{bhbg}
ds^{2}
&=&\left(\frac{U }{R }\right)^{3/2}\left(f(U) dt^{2} +d\vec{x}^2 +dx_4^2  \right)
+\left(\frac{R}{U }\right)^{3/2}\left( \frac{dU^2}{ f(U)} +U^2 d\Omega_4^2\right) \cr
e^\phi&=&g_s\left(\frac{U }{R }\right)^{3/4},\quad F_4 =\frac{2\pi N_c}{\Omega_4}\epsilon_4, \;\; f(U)= 1-\Big(\frac{U_{0}}{U}\Big)^{3}, \;\; R^3=\pi g_s N_c l_s^3.
\eea
There is a horizon at $U=U_0$, and the Hawking temperature and the horizon radius are related by \cite{Kruczenski:2003uq}
\be
U_0 =\frac{16 \pi^2}{9} R^3 T^2.
\ee
Introducing a dimensionless coordinate $\xi$  by $\frac{d\xi^2}{\xi^2}=\frac{dU^2}{U^2f(U)}$,
the background geometry becomes
\be\label{d4bhmetric}
ds^2 = \left(\frac{U }{R }\right)^{3/2}\left(f(U) dt^2 +d\vec{x}^2 + dx_4^{2} \right)
+\left(\frac{R}{U }\right)^{3/2}\left(\frac{U}{\xi}\right)^2\left(d\xi^2 +\xi^2 d\Omega_4^2\right),
\ee
where $U$ and $\xi$ are related by
\be
\left(\frac{U}{U_{0}}\right)^{3/2} = \frac{1}{2}\left(\xi^{3/2}+\frac{1}{\xi^{3/2}}\right), \;\; {\rm and } \;\;
f = \left(\frac{1-\xi^{-3}}{1+\xi^{-3}}\right)^2 \equiv \frac{\omega_{-}^2}{\omega_{+}^2}.
\ee
In this geometry, we rescale the coordinates such that the black hole horizon is located at $\xi =1$.
The induced metric of probe $Dq$ brane can be written as 
\bea\label{indDq}
ds_{Dq}^2&\!=&\!\left(\frac{1 }{R }\right)^{3/2}\xi^{3/2}\omega_{+}\left(\frac{\o_-^2}{\o_+^2}dt^2 +d\vec{x}_d^2  \right) \cr
&&~~~~~~~~~~+R^{3/2}\frac{\omega_{+}^{1/3}}{\xi^{3/2}}\left[\left(1+\dot{Y}^2\right)d\r^2 +\r^2 d\Omega_{q-d-1}^2\right].
\eea
The form of free energy of $D_q$ brane can be obtained from (\ref{ham02}) as follows
\bea\label{hamDq2}
{\cal F}_{Dq} = \t_q \int d\rho \sqrt{\frac{\o_-^2}{\o_+^{2/3}} (1+\dot{Y}^2)} \sqrt{\tilde{Q}^2 + \left(\xi^{3/4} \o_+ \right)^{d-1} \left(\frac{\rho^2 \o_+^{1/3}}{\xi^{3/4}}\right)^{n}},
\eea
where $n=q-d-1$. By substituting  the embedding solution 
 into  eq.(\ref{symE}), we can calculate symmetry energy in terms of density $Y_{\infty}$, 
the asymptotic value of probe brane, which is related to  quark mass and temperature by 
\be
Y_{\infty} =\frac{2\pi l_s^2}{U_0} m_q =\frac{9 l_s}{8\pi R^3} \cdot \frac{m_q}{T^2}.
\ee
Therefore, if we fix quark mass, large value of $Y_{\infty}$ corresponds to low temperature and small $Y_{\infty}$ to high  temperature. 
%The results are drawn in Figure \ref{fig:S2_D4bh}.

$\bullet$ $D6$ probe brane:\\
In this case, at low temperature(large value of $Y_{\infty}$),  the symmetry energy   $S_2 =0.5 \tilde{Q}^{1/2}$, 
which is same as confining case. 
At high temperature(small value of $Y_{\infty}$), symmetry energy increase linearly in density.
 We can understand this from eq.(\ref{symE}). The symmetry energy is integration 
 of a function from $\rho_{min}$ to infinity. In the case of nuclear matter system, $\rho_{min}$ is zero since the probe brane ends at the tip of baryon vertex. On the other hand, in deconfining phase, probe brane ends on black hole horizon  with  non-zero $\rho_{min}$. Therefore, embedding for small asymptotic value start near equator $\rho_{min} =1$.  If density is small($\tilde{Q} <<1$),   we can ignore $\tilde{Q}^2$ term in denominator and 
  symmetry energy is just linear in $\tilde{Q}$. However,  for low temperature,   $Y_{\infty}$ is small and $\rho_{min} \to 0$  and 
 therefore  $\tilde{Q}^2$ in the denominator contributes to the integral by scaling the variables, giving the power 
 behavior we observed above. 
 The result is described in the figure \ref{fig:S2_D4bh}(a). 
  
$\bullet$ $D4$ probe brane:\\
In the case of $D4$ probe brane, we get similar result in confining case which is lineally increase as density increase. 
See the figure \ref{fig:S2_D4bh}(b). 

\begin{figure}[!ht]
\begin{center}
\subfigure[]{\includegraphics[angle=0, width=0.45\textwidth]{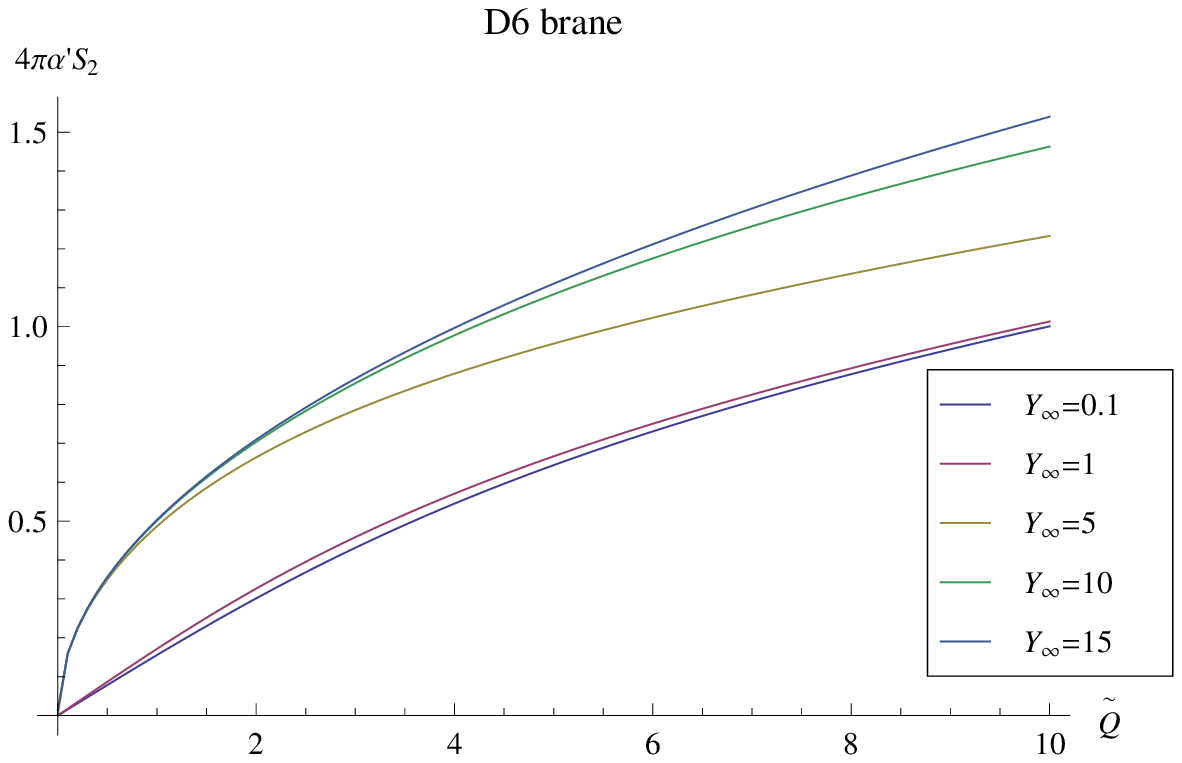}}
~~~~~
\subfigure[]{\includegraphics[angle=0, width=0.45\textwidth]{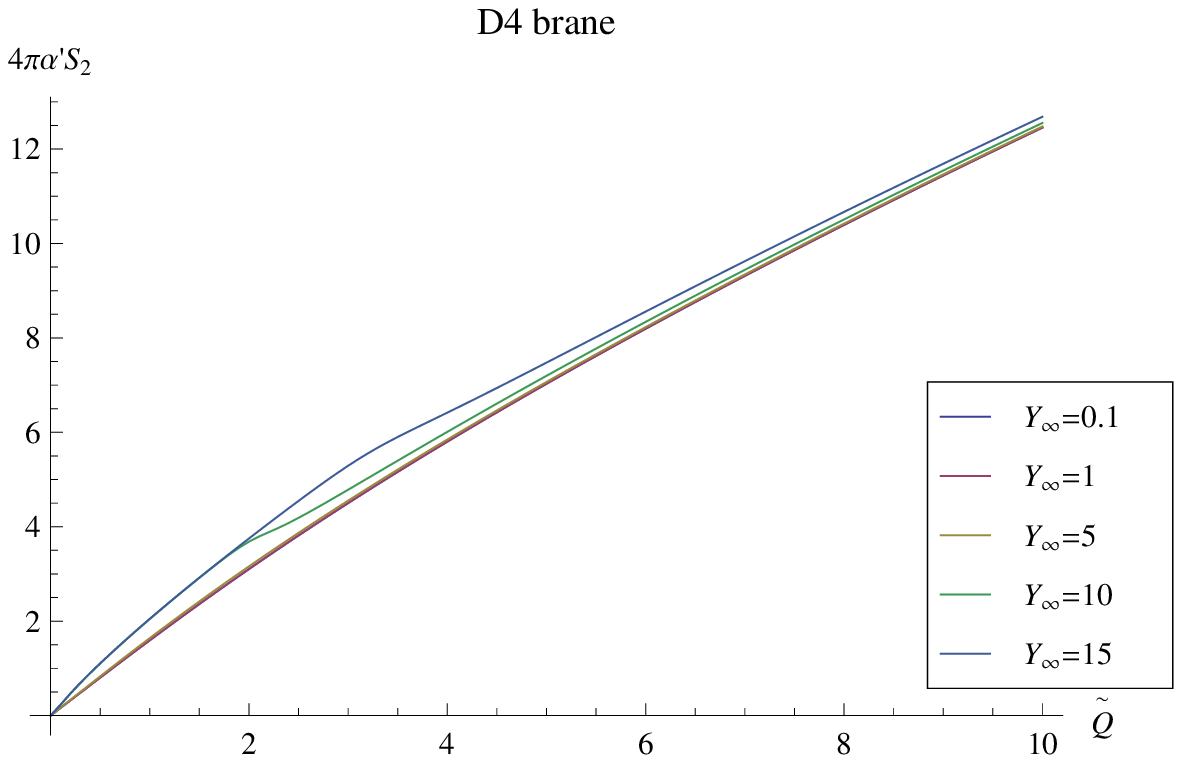}}
\caption{Density dependence of symmetry energy for several $Y_{\infty}$. \label{fig:S2_D4bh}}
\end{center}
\end{figure}

$\bullet$ $D2$ probe brane:\\
For $D2$ brane case symmetry energy become
\be
S_2 \sim \frac{\hat{Q}}{(\hat{Q}^2 +4)^{3/2}} \int d\rho \sqrt{\frac{\o_-^2}{\o_+^{2/3}} (1+\dot{Y}^2)},
\ee
which is also independent of details of embedding. And the result is almost same as Figure \ref{fig:S2_D4}(c).

\subsection{$D3$ brane background}
 We now  consider the black  $D3$ brane geometry;
\begin{eqnarray}
ds^{2}
&=&\frac{U^{2}}{R^{2}}\left(f(U) dt^{2}+d\vec{x}^{2}\right)
+R^{2}\left(\frac{dU^{2}}{f(U) U^{2}}+d\Omega_{5}^{2}\right),
\label{adsm}
%\\
%*dC_{(4)}&=&16\pi l_{s}^{4}N_{c}\epsilon_{(5)},
%\label{RR-flux}
\end{eqnarray}
where $R^{4}=2\lambda l_{s}^{4}$ and $f(U) =1-\left({U_{0}}/{U}\right)^{4}$.
% and $C_{(4)}$ is the background RR 4-form field. 
Here, $\lambda=g_{YM}^{2}N_{c}$ is the 't Hooft coupling of the YM theory.
There is a horizon at $U=U_{0}$, and the Hawking temperature is given by
\begin{eqnarray}
T=\frac{U_{0}}{\pi R^{2}}=\frac{U_{0}}{\sqrt{2\lambda}\pi l_{s}^{2}}.
\end{eqnarray}

Introducing a dimensionless coordinate $\xi$ defined by
$ {d\xi^2}/{\xi^2}= {dU^2}/({U^2f(U)})$, the bulk geometry becomes
\begin{eqnarray}
ds^{2}
&=&\frac{U^{2}}{R^{2}}\left(f dt^{2}+d\vec{x}^{2}\right)
+
\frac{R^{2}}{\xi^{2}}\left(d\xi^2 +\xi^2 d\Omega_5^2 \right), \label{metric}
\end{eqnarray}
where
$\xi^{2}\equiv y^{2}+\rho^{2}$ and $\rho$ is the radius of the 3-sphere. $U$ and $\xi$ are related by    
$ {U^{2}}/{U_0^2}=\frac{1}{2} (\xi^{2}+{1}/{\xi^{2}})\,$ and $f = ( {1- \xi ^{4}})^2/({1+ \xi^{4}} )^{2}.$
The induced metric on $Dq$ probe brane  and Hamiltonian density can be written as follows
\bea
ds_{Dq}^2 &=& \frac{\xi^2 \o_+}{R^2}\left(\frac{\o_-^2}{\o_+^2} dt^2 +d\vec{x}_d^2\right) +\frac{R^2}{\xi^2}\left[(1+\dot{Y}^2)d\rho^2 +\rho^2 d\O_{q-d-1}^2\right] \cr\cr
{\cal H}_{Dq} &=& \tau_q \sqrt{\frac{\o_-^2}{\o_+}(1+\dot{Y}^2)}\sqrt{\tilde{Q}^2 +\rho^{2n} \o_+^d}.
\eea
With completely  same method with the previous section, we  calculate
 symmetry energy for each probe brane($D7$, $D5$ and $D3$) in terms of density and asymptotic value of probe brane $Y_{\infty}$. In this case, the relationship beetween the asymptotic value of probe brane and temperature(or quark mass) is given by
\be
Y_{\infty}=\frac{2 m_q}{\sqrt{\lambda} T}.
\ee
The result is drawn in Figure \ref{fig:S2_D3bh}.
\begin{figure}[!ht]
\begin{center}
\subfigure[]{\includegraphics[angle=0, width=0.40\textwidth]{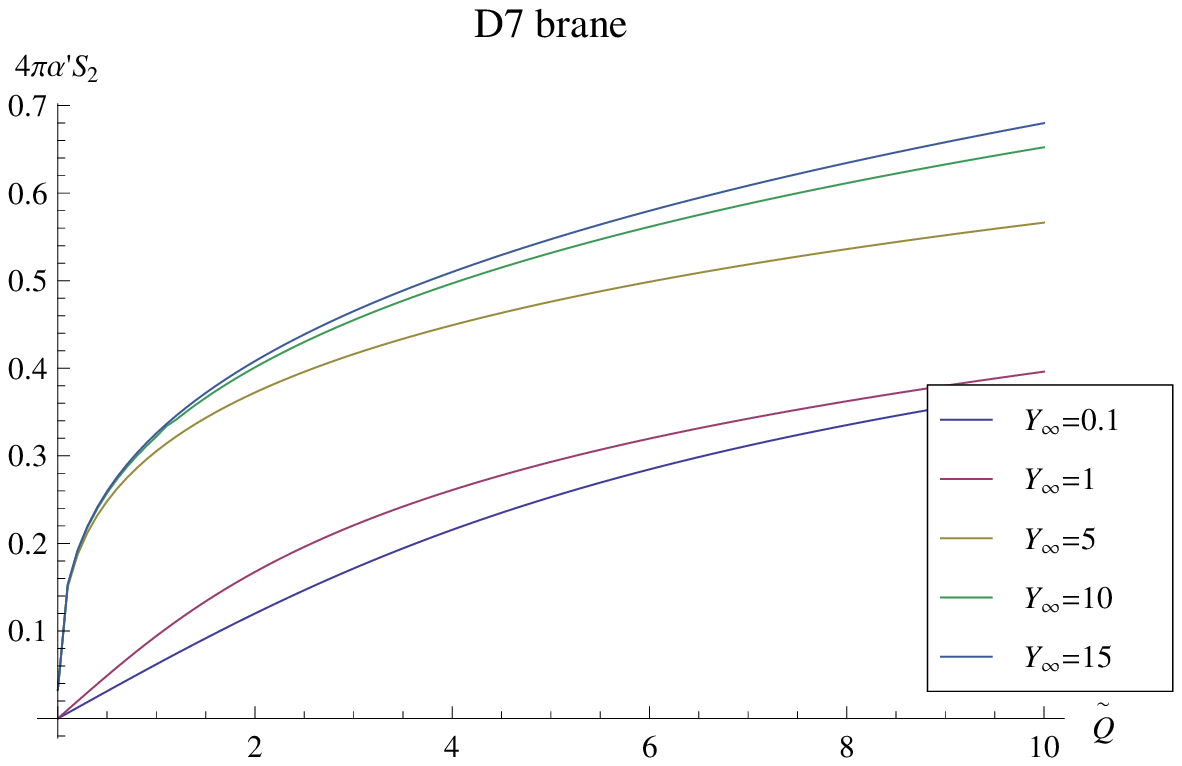}}
\subfigure[]{\includegraphics[angle=0, width=0.40\textwidth]{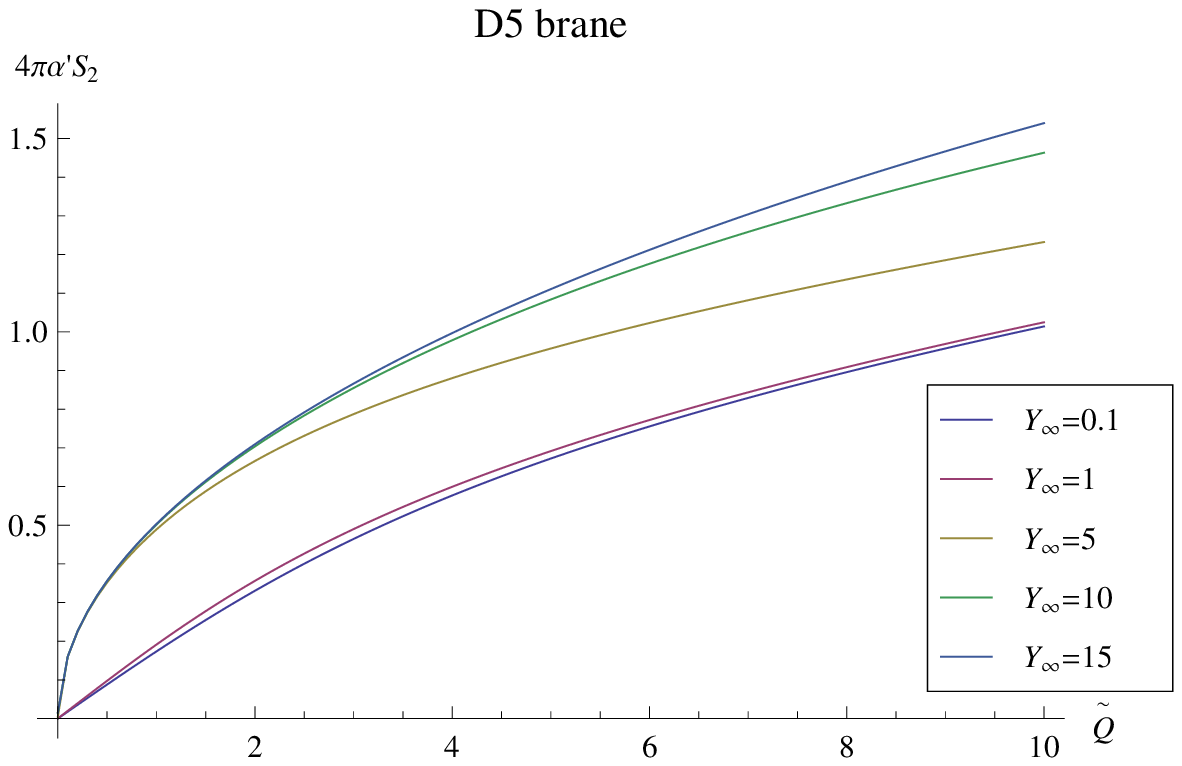}}
\subfigure[]{\includegraphics[angle=0, width=0.40\textwidth]{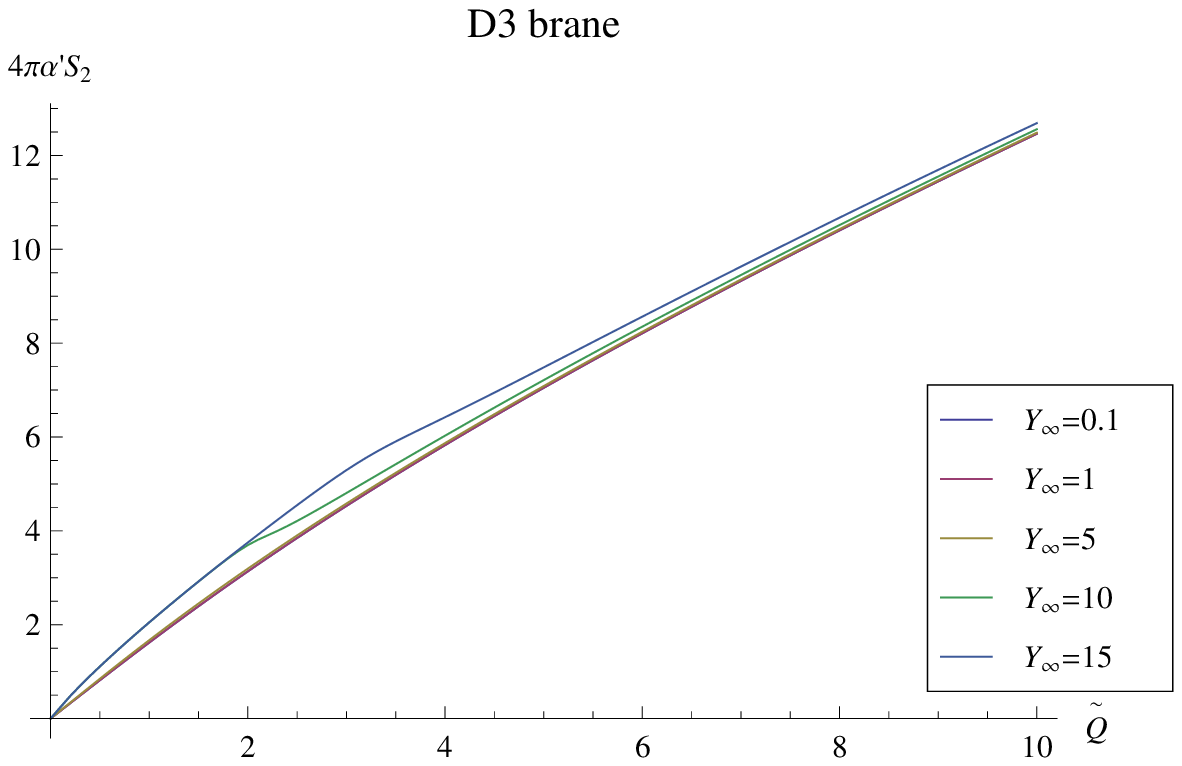}}
\caption{Density dependence of symmetry energy for several $Y_{\infty}$. \label{fig:S2_D3bh}}
\end{center}
\end{figure}
Same as the previous section,  we fix quark mass. Then, the asymptotic value of probe brane $Y_{\infty}$ proportional to the inverse temperature. In the case of $D7$ brane, for low temperature($Y_{\infty}=15$), the symmetry energy goes like  $S_2 = 0.32 \tilde{Q}^{1/3}$ while  at high temperature, the symmetry energy grows linearly. 
The symmetry energy line at low temperature for $D5$ probe case goes like $S_2 = 0.5 \tilde{Q}^{1/2}$ and it becomes linear at high temperature. In the case of $D3$ probe, symmetry energy has linear behavior for all temperature. 

%\newpage

\section{Scaling property and universality classes}
 In this section, we want to understand the scaling behavior of the symmetry energy 
 which is calculated in various models and various phases. 
 We first tabulate all the results  in table 1 and table 2. 
\begin{table}
\caption{$D4$ brane background}\label{S2D4}
\begin{tabular}{ |c|| c c c c c c c c c c | c | c | c || c |c|}
\hline
   & 0 & 1& 2 & 3 & 4 & 5 & 6 & 7 & 8 & 9 & $q$ & $d$ & $q-d-1$ & $S_2$ &$2\nu =n/d$ \\
\hline
$D4$&$\bullet$ &$\bullet$ &$\bullet$ &$\bullet$ &$\bullet$ & & & & & & & & && \\
\hline \hline 
$D2$&$\bullet$ &$\bullet$ & & & & $\bullet$ &&&&&2&1&0&${\cal O}(1)$ &-\\
\hline
$D4$&$\bullet$ &$\bullet$ &$\bullet$ & & & $\bullet$ &$\bullet$ &&&&4&2&1&$Q$& $1/2$\\
\hline
$D6$&$\bullet$ &$\bullet$ &$\bullet$ &$\bullet$ & & $\bullet$ &$\bullet$ &$\bullet$&&&6&3&2&$Q^{1/2}$ &$2/3$ \\
\hline
\end{tabular}
\end{table}
\begin{table}\label{S2D3}
\caption{$D3$ brane background}
\begin{tabular}{ |c|| c c c c c c c c c c | c | c | c || c |c|}
\hline
   & 0 & 1& 2 & 3 & 4 & 5 & 6 & 7 & 8 & 9 & $q$ & $d$ & $q-d-1$ & $S_2$ & $2\nu =n/d$  \\
\hline
$D3$&$\bullet$ &$\bullet$ &$\bullet$ &$\bullet$ & & & & & & & & & & &\\
\hline \hline 
$D3$&$\bullet$ &$\bullet$ &&  & $\bullet$ &$\bullet$ &&&&&3&1&1&$Q$&1 \\
\hline
$D5$&$\bullet$ &$\bullet$ &$\bullet$ &  & $\bullet$ &$\bullet$ &$\bullet$&&&&5&2&2&$Q^{1/2}$&1 \\
\hline
$D7$&$\bullet$ &$\bullet$ &$\bullet$ &$\bullet$ & $\bullet$& $\bullet$ &$\bullet$ &$\bullet$&&&7&3&3&$Q^{1/3}$&1 \\
\hline
\end{tabular}
\end{table}
We emphasize that the result is based on the actual numerical calculation without any approximation. 

We now re-derive   these power behavior of symmetry energy  analytically. 
To do this, we consider the ideally simplified case: BPS metric background and flat embedding of probe branes. 
In this case,  background geometry becomes geometry of black $Dq$ brane;
\be\label{Dpmetric}
ds_{10}^2 = Z_p^{-1/2}(-dt^2 +d\vec{x}_p^2) +Z_p^{1/2} d\vec{x}_{\perp}^2,
\ee
with $e^{2\phi} = Z_p^{\frac{3-p}{2}}.$
We now consider the general result for the symmetry energy 
\be
S_2= 2\tau_q \int d\r \frac{\tilde{Q}\sqrt{G_{tt}G_{\r\r}}e^{-2\phi}G_{xx}^d G_{\O\O}^{n}}{\left(\tilde{Q}^2 +4 e^{-2\phi}G_{xx}^d G_{\O\O}^{n}\right)^{3/2}}.
\ee
which is already given in eq. (\ref{symE}).
Then, the term in the square root becomes 1 and  
\bea
e^{-2\phi} G_{xx}^d \left(\frac{G_{\perp\perp}}{\xi^2}\right)^{q-d-1} 
&=& Z_p^{\frac{p-3}{2}} \cdot Z_p^{-\frac{d}{2}}\cdot Z_p^{\frac{q-d-1}{2}}\cr
&=& Z_p^{\frac{1}{2}(p+q-2d-4)}.
\eea
The value of $p+q-2d$ is precisely equal to the number of Neuman-Dirichlet (ND) direction  of $Dp/Dq$ system. Therefore, if we focus on the system which is supersymmetric configurations or  a smooth deformation of them,  the exponent becomes zero.  In this case 
the symmetry energy (\ref{symE}) for flat embedding can be calculated analytically: 
\bea\label{flatS2}
S_2 & =& 2\tau_q \int d\r \frac{\tilde{Q} \r^{2 n}}{\left(\tilde{Q}^2 + 4\r^{2 n}\right)^{3/2}}
=c_n\tilde{Q}^{\frac{1}{n}},
\eea
where $c_n= 2\tau_q\frac{2^{-2-1/n}\Gamma\left(\frac{1}{2n}\right)\Gamma\left(\frac{n-1}{2n}\right)}{n^2 \sqrt{\pi}}$ and $n=q-d-1$ so that the density dependence of symmetry energy is
$ S_2 \sim Q^{\frac{1}{q-d-1}}$. 
This result reproduces all the result we obtained in previous section numerically. 

Notice that both  the background and embedding used here  are far from the real situation: 
real background is a deformation of such BPS solution 
and the embedding is non-trivial deformation from such a flat embedding. 
Nevertheless, the scaling exponent  is the same as the actual configuration used for numerical computation of previous sections. 
The point  is that neither  smooth deformation  of the metric nor the deformation of the 
embedding shape seem to  change the scaling behavior of the symmetry energy. 
The exponent of symmetry energy  depends only  on the dimensionality of 
probe brane and dimension of non-compact directions. 
{\it Therefore the scaling exponents depend only on the universality classes. }

\section{Discussion}
In this paper, we calculated the asymmetry energy for  both nuclear matter as well as the quark matter. 
The symmetry energy has a power like density dependence with characteristic exponent 
which is invariant under the smooth deformation of the metric as well as smooth deformation of the 
embedding. It only depends on the dimensionality of the D-brane system modeling the QCD dynamics. 
Therefore it is a index for the universality class. 
The physical interpretation of the scaling exponent is still open question but we give a trial interpretation 
below. 

%\noindent
{\it Discussion: Non-fermi liquid nature of the nuclear matter. } %Here we try to understand the implication of the scaling exponent.
According to the fermi gas model of nuclei,
the Energy of the nuclei is the sum of the kinetic energies 
of neutrons and protons which are moving non-relativistically, namely, 
\bea
E&=&\frac{3}{5} (N_p \epsilon_{Fp}+ N_n \epsilon_{Fn})\\
&=& \frac{3}{5}  \epsilon_{F} A +\frac1{3}\epsilon_F A\delta^2 +{\cal O}(\delta^4),
\eea
where $A=N_p +N_n$ is the mass number and $\epsilon_F$ is the fermi energy for $A$, 
$\delta=\frac{(N_p -N_n)}{(N_p +N_n)}$.
One can see that the symmetry energy per nucleon is $\epsilon_F/3$.
Therefore fermi gas model demonstrates the origin of the symmetry energy as the Pauli principle.
What happens if we include the interaction energy of the nucleons? The answer is largely unknown. Depending on how one includes the interaction, the answers are different from one another. In some of the traditional approach, interactions are taken care of by  adding a polynomial in density, which means that scaling property ($\sim \rho^{\alpha}$) should be destroyed by the interaction.  However, this is not what we get. In our approach,  scaling property remains even in the  large density limit. 
Notice that for the most interesting case of $D4$/$D6$ and $D3$/$D7$, the scaling exponents are $1/2$ and $1/3$ respectively. 
Comparing this with the non-interacting gas which gives $S_2 \sim \rho^{2/3}$, one can see that 
the symmetry energy of $D4$/$D6$ implies an anomalous dispersion relations which is neither relativistic nor non-relativistic one.  For the $D3$/$D7$ case, the scaling is the same as the 
relativistic particles. 
If we naively extrapolate the relation $S_2 \sim \epsilon_F$, 
$S_2\sim \rho^{1/2}$ means the anomalous dispersion relation $\epsilon\sim k^{3/2}$. 
One may want to associate this as the example of transformation of a particle 
to  un-particle  \cite{Georgi:2010zz}

One may also expect that such anomalous  dispersion relation is related to the non-fermi liquid nature of the 
strongly interacting fermion system. 
In the presence of the fermi sea, one expect that elementary excitations are quasi particles with renormalized charge and mass. However, when the interactions are strong, such quasi-particles will lose applicability. 
Using the AdS/CFT duality and utilizing the AdS at UV and 
the $AdS_2$  at the IR, it was shown in\cite{Faulkner:2009wj} that 
\be
G(\omega,k)=\frac{1}{\omega-v_F k_\perp-C\omega^{2\nu}} 
,\ee
where $k_\perp$ is the momentum measured from the fermi surface. 
If $\nu <1/2$, the dispersion relation becomes $\omega\sim k_\perp^{1/2\nu}$. For example, 
if $\nu=1/3$ we get $\omega\sim k_\perp^{3/2}$. 
In our case, the situation is more subtle since the fermi surface is not stable under 
turning on temperature and fermi sea is not so calm: 
the low lying states below `fermi sea' can be all excited so that it should be called fermi ball \cite{Lee:2008xf} rather than fermi surface, 
and $k_\perp$ should be replace by  $ k$. 
These phenomena are all beyond the fermi liquid behavior.  For D3/D7 case, the exponent indicate that 
the system  is marginally fermi liquid case. 
More systematic investigation on this matter is strongly desired.

\section*{Acknowledgments}
This work was supported by the  NRF grant funded by the Korea government(MEST) through the  Mid-career Researcher Program  with grant No. 2010-0008456, 
and it is also supported by NRF through SRC program Center for Quantum Space-time  with grant number 2005-0049409.  

\appendix
\section{Free energy of $Dq$ brane}\label{AppA}
We start from 10 dimensional metric for $Dp$ brane with Lorenzian signature as
\bea\label{10dmetric}
ds_{10}^2 &=&-G_{tt} dt^2 +G_{xx}d\vec{x}_{p}^2+G_{rr}d r^2 + G_{\perp \perp} d\O_{\perp}^2 \cr
&=& -G_{tt} dt^2 +G_{xx}d\vec{x}_{p}^2+G_{\perp \perp} \left(\frac{G_{rr}}{G_{\perp \perp} }d r^2 + d\O_{\perp}^2\right),
\eea
where $d\O_{\perp}^2$ is metric of $S^{(8-p)}$. To make the orthogonal space to $Dp$ brane, we introduce new coordinate;
\be
\sqrt{\frac{G_{rr}}{G_{\perp \perp}}}dr =\frac{d\xi}{\xi},
\ee
then, the metric (\ref{10dmetric}) becomes
\be
ds_{10}^2 = -G_{tt} dt^2 +G_{xx}d\vec{x}_{p}^2+\frac{G_{\perp \perp}}{\xi^2}\left(d\xi^2 +\xi^2 d\O_{\perp}^2 \right).
\ee
Now, we put probe $Dq$ brane in this background with $d$ dimensional uncompact direction. We can decompose transverse direction into parallel and perpendicular to the probe brane as follows
\bea\label{10dmetric2}
ds_{10}^2 &=&-G_{tt} dt^2 +G_{xx}(d\vec{x}_{d}^2+d\vec{x}_{p-d}^2)+\frac{G_{\perp \perp}}{\xi^2}\left(dR_{q-d}^2 +dR_{9-p-q+d}^2 \right)\cr
&=& -G_{tt} dt^2 +G_{xx}(d\vec{x}_{d}^2+d\vec{x}_{p-d}^2) \cr
&&+\frac{G_{\perp \perp}}{\xi^2}\left(d\rho^2 +\rho^2 d\O_{q-d-1}^2 +dY^2 +Y^2 d\O_{8-p-q+d}\right).
\eea
We assume that only $Y$ coordinate has $\rho$ dependence and the other coordinates which perpendicular to probe brane is constant, then the induce metric on probe brane can be written as
\bea\label{inducedg}
ds_{Dq}^2 &=& -G_{tt} dt^2 +G_{xx} d\vec{x}_{d}^2 +\frac{G_{\perp \perp}}{\xi^2}(1+\dot{Y}^2)d\rho^2 +\frac{G_{\perp \perp}}{\xi^2}\rho^2 d\O_{q-d-1}^2 \cr
&\equiv& -G_{tt} dt^2 +G_{xx} d\vec{x}_{d}^2 +G_{\rho\rho}d\rho^2 + G_{\O\O} d\O_{q-d-1}^2,
\eea
where $\dot{Y} =\partial Y/\partial \rho$.\par
To introduce number density in boundary theory, we turn on time component of $U(1)$ gauge field $A_t (\rho)$ on the probe brane and set all the other components to be zero. Then we can write DBI action as follows;
\bea
S_{Dq} &=& -\mu_{q} \int d\sigma^{q+1} e^{-\phi} \sqrt{{\rm det}(g+2\pi \a' F)}\cr
&=& -\mu_{q} \int dt \,d^{d}x \, d\O_{q-d-1}\, d\rho\, e^{-\phi}\sqrt{G_{xx}^d G_{\O\O}^{q-d-1}(G_{tt}G_{\r\r} -\tilde{F}^2)} \cr
&\equiv& \int dt\,d\r\, {\cal L}_{Dq},
\eea
where $\tilde{F} \equiv =2\pi\a' F_{t\r} =2\pi\a' \partial_\r A_t (\r)$.\par
From the equation of motion for gauge field, we can define conserved charge;
\bea
\frac{\partial {\cal L}_{Dq}}{\partial \tilde F} =\frac{\tau_q e^{-\phi} G_{xx}^d G_{\O\O}^{q-d-1}\tilde{F}}{\sqrt{G_{xx}^d G_{\O\O}^{q-d-1}(G_{tt}G_{\r\r} -\tilde{F}^2)}}\equiv \frac{Q}{2\pi\a'},
\eea
where $\tau_q = \mu_q V_d \O_{q-d-1}$. Then we can get the relation between field strength and charge as
\be
\tilde{F} =\frac{\tilde{Q}\sqrt{G_{tt}G_{\r\r}}}{\sqrt{\tilde{Q}^2 +e^{-2\phi}G_{xx}^d G_{\O\O}^{q-d-q}}},
\ee
where $\tilde{Q} =Q/2\pi \a'\tau_q$.\par
After Legengdre transformation for Lagrangian density, we can get free energy of $D_q$ brane;
\bea
{\cal F}_{Dq} &=& \int d\rho \left(\tilde{F}\frac{\partial {\cal L}_{Dq}}{\partial \tilde F} -{\cal L}_{Dq}\right) \cr
&=& \tau_q \int d\rho \sqrt{G_{tt}G_{\r\r}}\sqrt{\tilde{Q}^2 +e^{-2\phi} G_{xx}^d G_{\O\O}^{q-d-1}}\label{ham01}\\
&=&\tau_q \int d\rho \sqrt{G_{tt}\frac{G_{\perp \perp}}{\xi^2}(1+\dot{Y}^2)}\sqrt{\tilde{Q}^2 +e^{-2\phi}\, G_{xx}^d \left(\frac{G_{\perp \perp}}{\xi^2}\rho^2 \right)^{q-d-1}}\label{ham02}.
\eea

\section{Baryon vertex and force balance condition}\label{AppB}
In this work, we consider $D3$ or $D4$ brane as a background. In $D3$ brane background, $D5$ brane wrapping  on $S^5$ with $N_c$ fundamental strings is interpreted as baryon vertex and a spherical $D4$ brane on $S^4$ is baryon vertex in $D4$ brane background. In the both case, background metric can be written in general form;
\bea
ds_{10}^2 &=& -G_{tt} dt^2 +G_{xx}d\vec{x}_{p}^2+\frac{G_{\perp \perp}}{\xi^2}\left(d\xi^2 +\xi^2 d\O_{8-p}^2 \right)\cr
&=& -G_{tt} dt^2 +G_{xx}d\vec{x}_{p}^2+\frac{G_{\perp \perp}}{\xi^2}d\xi^2 + G_{\perp \perp}\left(d\theta^2 +\sin^2\theta d\O_{7-p}^2\right),
\eea
here, we assume that the $D(8-p)$ brane has $SO(7-p)$ symmetry, then $\xi$ depends on polar angle $\theta$ only. Then the induced metric on $D(8-p)$ brane is
\be\label{induceBV}
ds_{BV}^2 = -G_{tt} dt^2 +\frac{G_{\perp \perp}}{\xi^2}\left(\xi^2 +\xi'^2 \right)d\theta^2 + G_{\perp \perp}\sin^2\theta d\O_{7-p}^2.
\ee
The DBI action for $D(8-p)$ brane can be written as followes;
\be\label{SBV}
S_{BV} = -\mu_{8-p} \int d^{9-p} \sigma e^{-\phi} \sqrt{{\rm det}(g+2\pi\a' F)} +\mu_{8-p} \int 2\pi\a' A_t \wedge G_{(7-p)},
\ee
where $\mu_{8-p}$ is a D-brane tension and $G_{(7-p)}$ is Ramon-Ramon field which couples to the original $Dp$ brane. After substituting induce metric (\ref{induceBV}), we can get
\bea\label{LBV}
S_{BV} &=& \tau_{8-p} \int dt d\theta \sin^{(7-p)}\theta \left[-R^{(p-7)} e^{-\phi} G_{\perp \perp}^{\frac{7-p}{2}}\sqrt{\frac{G_{tt} G_{\perp \perp}}{\xi^2} \left(\xi^2 +\xi'^2\right) -\tilde{F}^2} +(7-p)\tilde{A}_t \right] \cr
&\equiv& \int dt d\theta {\cal L}_{BV},
\eea
where
\be
\tau_{8-p} = \mu_{8-p} \Omega_{7-p} R^{7-p},~~~~\tilde{A}_t = 2\pi\a' A_t.
\ee
The displacement can be obtained by derivative the action with respect to $\tilde{F}$,
\be
\frac{\partial {\cal L}_{BV}}{\partial \tilde{F}} =\frac{\tau_{8-p} R^{p-7} \sin^{7-p}\theta e^{-\phi}G_{\perp \perp}^{\frac{7-p}{2}}\cdot \tilde{F}}{\sqrt{\frac{G_{tt} G_{\perp \perp}}{\xi^2} \left(\xi^2 +\xi'^2\right) -\tilde{F}^2}} \equiv -D(\theta),
\ee
and the equation of motion for gauge field is given by
\be
\partial_{\theta} \tilde{D}(\theta) =-(7-p) \sin^{7-p}\theta,
\ee
where $\tilde{D}(\theta) \equiv D(\theta)/\tau_{8-p}$. By integrating, we get the solution in terms of hypergeometric function as follows
\be
\tilde{D}(\theta)=c_0 + (7-p) \cos\theta \,\, _2F_1\left(\frac{1}{2},\frac{p-6}{2},\frac{3}{2},\cos^2\theta\right).
\ee
For example, in the case of $D4$ brane $(p=4)$, we get
\be
\tilde{D}(\theta)=c_0 +\cos\theta(3-\cos^2\theta),
\ee
which is same as result in \cite{Seo:2008qc}. The integration constant $c_0$ is determined such that $\tilde{D}$ vanishes at $\theta=0$ which means all fundamental strings are attached on north pole of $D(8-p)$ brane. After substituting the solution of equation of motion, we can rewrite DBI action in terms of displacement which is called as 'Hamilotnian' because the procedure to get it is similar to Legendre transformation.
\be\label{HBV}
{\cal H}_{BV} =\tau_{8-p} \sqrt{\frac{G_{tt} G_{\perp \perp}}{\xi^2} \left(\xi^2 +\xi'^2\right)}\sqrt{\tilde{D}^2 + R^{2(p-7)}\sin^{2(7-p)}\theta\,\,e^{-2\phi}G_{\perp\perp}^{7-p}}.
\ee
As discussed in previous works, we can calculate force at the cusp of single $D(8-p)$ brane as follows,
\bea
F_{BV} &=& \frac{\delta {\cal H}_{BV}}{\delta \xi_c} \Bigg|_{\rm on-shell} 
= \frac{\partial {\cal H}_{BV}}{\partial \xi'}\Bigg|_{\theta=\pi}\cr\cr
&=& \tau_{8-p} \frac{(7-p)\sqrt{\pi}\Gamma\left(\frac{8-p}{2}\right)}{2 \cdot\Gamma\left(\frac{9-p}{2}\right)} \sqrt{\frac{G_{tt} G_{\perp \perp}}{\xi_c^2}}\frac{\xi_c'}{\sqrt{\xi_c^2 +\xi_c'^2}}\cr\cr
&=& \frac{N_c}{2\pi \a'} \sqrt{\frac{G_{tt} G_{\perp \perp}}{\xi_c^2}}\cdot\frac{\xi_c'}{\sqrt{\xi_c^2 +\xi_c'^2}},
\eea
where $\xi_c$ and $\xi_c'$ denote to the value of $\xi$ and it's derivative at the cusp. The overall factor becomes $N_c/2\pi\a'$ for both of $D3$ and $D4$ case.

Now, we can get boundary condition for probe $Dq$ brane by imposing {\it 'force balance condition'}. The force at the cusp of probe brane can be obtained from Hamiltonian of probe brane  (\ref{ham02});
\bea
F_{Dq} &=& \frac{\delta {\cal H}_{Dq}}{\delta Y_c} \Bigg|_{\rm on-shell} 
= \frac{\partial {\cal H}_{Dq}}{\partial \dot{Y}}\Bigg|_{\rho=0}\cr\cr
&=& \frac{Q}{2\pi\a'} \sqrt{\frac{G_{tt} G_{\perp \perp}}{\xi_c^2}} \frac{\dot{Y_c}}{\sqrt{1+\dot{Y_c}^2}},
\eea
where $Y_c$ and $\dot{Y}_c$ denote to the value of $Y$ and it's derivative at the cusp of probe brane. From the force balance condition
\be
F_{Dq}(Q) = N_B\cdot F_{BV} =\frac{Q}{N_c} F_{BV},
\ee
we can get the boundary condition for probe brane
\be\label{bc}
\dot{Y}_c = \frac{ \xi_c'}{Y_c}
\ee
which does not depends on type of original brane or probe brane.

\end{document}